\newif\ifpdflatex    
\pdflatextrue           

\documentclass[fleqn,usenatbib,useAMS,twocolumn]{aastex631}
\usepackage{amsmath,esint}
\usepackage{url,aasmacros}
\usepackage{natbib}
\usepackage{soul, CJK}
\usepackage{multirow}
\usepackage{tablefootnote}

\usepackage{bm}
\usepackage{xspace}
\usepackage{color}
\usepackage{enumitem}
\usepackage{booktabs}

\usepackage{graphicx}	
\usepackage{amsmath}	
\usepackage{amssymb}	
\usepackage{bm}		
\usepackage{url}
\usepackage{amsbsy}
\usepackage{xspace}
\usepackage{hyperref}
\usepackage{longtable}
\usepackage[colorinlistoftodos]{todonotes}
\usepackage[flushleft]{threeparttable}
\usepackage{longtable}

\usepackage[T1]{fontenc}
\usepackage{ae,aecompl}

\def\lesssim{\mathrel{\hbox{\rlap{\hbox{\lower5pt\hbox{$\sim$}}}\hbox{$<$}}}}
\def\gtrsim{\mathrel{\hbox{\rlap{\hbox{\lower5pt\hbox{$\sim$}}}\hbox{$>$}}}}

\usepackage{upgreek}                                                  
\usepackage{xspace}                                                       
\newcommand{\um}{$\upmu$m\xspace}            
\def\apjl{ApJ}%
\def\apjs{ApJS}%
\def\aap{A\&A}%
\shorttitle{RCB stars from PGIR}
\shortauthors{Karambelkar et al.}

\begin{document}
\title{An infrared census of R Coronae Borealis Stars II -- Spectroscopic classifications and implications for the rate of low-mass white dwarf mergers}
\author[0000-0003-2758-159X]{Viraj R. Karambelkar}
\email{viraj@astro.caltech.edu}
\affiliation{Cahill Center for Astrophysics, California Institute of Technology, Pasadena, CA 91125, USA}

\author{Mansi M. Kasliwal}
\affiliation{Cahill Center for Astrophysics, California Institute of Technology, Pasadena, CA 91125, USA}

\author{Patrick Tisserand}
\affiliation{Sorbonne Universit\'e, CNRS, UMR 7095, Institut d’Astrophysique de Paris, 98 bis bd Arago, 75014 Paris, France}

\author[0000-0003-3768-7515]{Shreya Anand}
\affiliation{Cahill Center for Astrophysics, California Institute of Technology, Pasadena, CA 91125, USA}

\author{Michael C. B. Ashley}
\affiliation{School of Physics, University of New South Wales, Sydney, NSW 2052, Australia}

\author[0000-0001-8038-6836]{Lars Bildsten}
\affiliation{Department of Physics, University of California, Santa Barbara, CA 93106, USA}
\affiliation{Kavli Institute for Theoretical Physics, University of California, Santa Barbara, CA 93106, USA}

\author[0000-0002-0141-7436]{Geoffrey C. Clayton}
\affiliation{Department of Physics \& Astronomy, Louisiana State University, Baton Rouge, LA 70803, USA}
\affiliation{Space Science Institute,
4765 Walnut St, Suite B
Boulder, CO 80301, USA}

\author{Courtney C. Crawford}
\affiliation{Sydney Institute for Astronomy (SIfA), School of Physics, University of Sydney, Sydney, NSW 2006, Australia}

\author[0000-0002-8989-0542]{Kishalay De}
\altaffiliation{NASA Einstein Fellow}
\affil{MIT-Kavli Institute for Astrophysics and Space Research, 77 Massachusetts Ave., Cambridge, MA 02139, USA}


\author{Nicholas Earley}
\affiliation{Cahill Center for Astrophysics, California Institute of Technology, Pasadena, CA 91125, USA}

\author[0000-0001-9315-8437]{Matthew J. Hankins}
\affil{Arkansas Tech University, Russellville, AR 72801, USA}

\author{Xander Hall}
\affiliation{Cahill Center for Astrophysics, California Institute of Technology, Pasadena, CA 91125, USA}

\author[0000-0001-8740-0127]{Astrid Lamberts}
\affil{Laboratoire Lagrange, Universit\'e Cote d’Azur, Observatoire de la Cote d’Azur, CNRS, Bd de l’Observatoire, 06300 Nice, France}
\affil{Laboratoire Artemis, Universit\'e Cote d’Azur, Observatoire de la Cote d’Azur, CNRS, Bd de l’Observatoire, 06300 Nice, France}

\author{Ryan M. Lau}
\affil{NSF's National Optical-Infrared Astronomy Research Laboratory, 950 N. Cherry Ave., Tucson, AZ 85719, USA}

\author{Dan McKenna}
\affil{Caltech Optical Observatories, California Institute of Technology, Pasadena, CA 91125, USA}

\author{Anna Moore}
\affil{Research School of Astronomy and Astrophysics, Australian National University, Canberra, ACT 2611, Australia}

\author{Eran O. Ofek}
\affil{Department of Particle Physics and Astrophysics, Weizmann Institute of Science, Rehovot 76100, Israel}

\author[0000-0001-7062-9726]{Roger M. Smith}
\affil{Caltech Optical Observatories, California Institute of Technology, Pasadena, CA 91125, USA}

\author[0000-0002-4622-796X]{Roberto Soria}
\affil{College of Astronomy and Space Sciences, University of the Chinese Academy of Sciences, Beijing 100049, China}
\affil{INAF-Osservatorio Astrofisico di Torino, Strada Osservatorio 20, I-10025 Pino Torinese, Italy}
\affil{Sydney Institute for Astronomy, School of Physics A28, The University of Sydney, Sydney, NSW 2006, Australia}

\author[0000-0001-9226-4043]{Jamie Soon}
\affil{Research School of Astronomy and Astrophysics, Australian National University, Canberra, ACT 2611, Australia}

\author[0000-0001-9304-6718]{Tony Travouillon}
\affil{Research School of Astronomy and Astrophysics, Australian National University, Canberra, ACT 2611, Australia}


\begin{abstract}
    We present results from a systematic infrared (IR) census of R Coronae Borealis (RCB) stars in the Milky Way, using data from the Palomar Gattini IR (PGIR) survey. R Coronae Borealis stars are dusty, erratic variable stars presumably formed from the merger of a He-core and a CO-core white dwarf (WD). PGIR is a 30\,cm $J$-band telescope with a 25\,deg$^{2}$ camera that surveys 18000\,deg$^{2}$ of the northern sky ($\delta>-28^{o}$) at a cadence of 2\,days. Using PGIR \emph{J-}band lightcurves for $\sim$60 million stars together with mid-IR colors from WISE, we selected a sample of 530 candidate RCB stars. We obtained near-IR spectra for these candidates and identified 53 RCB stars in our sample. Accounting for our selection criteria, we find that there are a total of $\approx350^{+150}_{-100}$ RCB stars in the Milky Way. Assuming typical RCB lifetimes, this corresponds to an RCB formation rate of 0.8 -- 5\,$\times$\,10$^{-3}$\,yr$^{-1}$, consistent with observational and theoretical estimates of the He-CO WD merger rate. We searched for quasi-periodic pulsations in the PGIR lightcurves of RCB stars and present pulsation periods for 16 RCB stars. We also examined high-cadenced \emph{TESS} lightcurves for RCB and the chemically similar, but dustless hydrogen-deficient carbon (dLHdC) stars. We find that dLHdC stars show variations on timescales shorter than RCB stars, suggesting that they may have lower masses than RCB stars. Finally, we identified 3 new spectroscopically confirmed and 12 candidate Galactic DY\,Per type stars -- believed to be colder cousins of RCB star -- doubling the sample of Galactic DY\,Per type stars.\\ \\ \\ \\
\end{abstract}

\section{Introduction}
\label{sec:intro}
R Coronae Borealis (RCB) stars are an enigmatic class of stellar variables, notable for extreme brightness variations and peculiar chemical compositions \citep{ Clayton12}. They show deep, rapid declines in their optical brightness ($\lesssim$ 9 mag in $V$ band), which can last hundreds of days before rising back to their initial state \citep{Clayton1996}. In addition, they have helium-rich atmospheres with an acute deficiency of hydrogen and an overabundance of carbon \citep{Asplund2000}. The chemical compositions of RCB stars point to them being remnants of white dwarf (WD) mergers between a He-WD and a CO-WD \citep{Webbink84,Clayton2005, Clayton2007}, possibly making them low-mass analogs of type Ia supernova progenitors \citep{Fryer2008}. Furthermore, close He-CO WD binaries are expected to be the dominant population of gravitational wave sources detected by LISA \citep{Lamberts2019}. RCB stars represent the fate of this population and can potentially be used to infer the merger rates of these.

RCB stars have several intriguing properties that present challenges to models of stellar evolution. Their photometric declines are the result of mass-loss episodes that produce dust and obscure the star \citep{Clayton1992}. The origin of these mass-loss episodes is still not known. While at maximum light, some RCB stars pulsate with periods between 40--100 days \citep{Lawson1997}, with the pulsation periods likely depending on their effective temperatures (T$_{\rm{eff}}$) which range from 4000 -- 8000\,K \citep{Tisserand2023a, Crawford2023}. The origin of these pulsations is still not known -- they have been attributed either to the strange-mode instability \citep{Saio2008, Gautschy2023}, or thought to be solar-like oscillations in helium-rich envelopes \citep{Wong2024}. 

RCB stars are closely related to the class of dustless Hydrogen-deficient Carbon stars (dLHdC stars, \citealt{Warner1967, Tisserand2022}). dLHdC stars have similar chemical compositions to RCB stars, but show no signs of dust-formation. Together, RCB and dLHdC stars constitute the class of Hydrogen-deficient Carbon (HdC) stars. The differences between RCB stars and dLHdC stars have been explored only recently, based on their positions in the HR diagram \citep{Tisserand2022}, their oxygen isotope ratios \citep{Karambelkar2022}, their strontium-abundances \citep{Crawford2022} and their H and Li content \citep{Crawford2023}. These initial results suggest that RCB stars may be more massive than dLHdC stars. These results are based on small samples of these stars, and it still remains a mystery as to why RCB stars form dust, while dLHdC stars do not. 

A third class of stars called DY\,Per type stars is thought to be a colder sub-class of RCB stars (with T$_{\rm{eff}} \approx 3500$\,K), marked by shallower and more symmetric declines in their lightcurves \citep{Alcock2001}. Only 3 DY\,Per type stars have been confirmed in the Milky Way \citep{Tisserand2008, Tisserand2013}, while about a dozen have been confirmed in the Magellanic Clouds \citep{Alcock2001,Tisserand2004,Tisserand2009}. These stars have a hydrogen-deficiency, a high carbon-abundance \citep{Zacs2007} and have been recently found to have a high $^{18}$O abundance \citep{Bhowmick2018, Garcia-Hernandez2023} -- resembling RCB stars. On the other hand, their luminosities and infrared colors are very similar to cool N-type carbon stars and differ significantly from RCB stars \citep{Alcock2001, Soszynski2009, Tisserand2009}. It is debated whether DY\,Per type stars are colder RCB stars originating in WD mergers, or classical carbon stars undergoing strong dust-formation. 

When RCB stars are enshrouded by dust, they appear brighter at infrared wavelengths than the optical \citep{Feast1997b}. The photometric declines are also shallower in the infrared ($\lesssim$ 3 mag in \emph{J-} band) than in the optical \citep{Karambelkar2021}. Despite this, most previous searches for RCB stars have focused on optical surveys such as MACHO (\citealt{Alcock2001,Zaniewski2005}), OGLE \citep{Tisserand2011}, EROS-2 \citep{Tisserand2004,Tisserand2008,Tisserand2009}, the Catalina Survey \citep{Lee2015}, All Sky Automated Survey for Supernovae (ASASSN, \citealt{Tisserand2013,Shields2019,Otero2014}), the Palomar Transient Facility \citep{Tang2013} and the Zwicky Transient Facility \citep{Lee2020}. 
 
The first IR-search for RCB stars was carried out by \citet{Tisserand2012}, who used WISE and 2MASS colors to publish a list of RCB candidates. \citet{Tisserand2020} subsequently refined the selection criteria and published a catalog of 2356 RCB candidates and 40 new RCB stars confirmed via optical spectroscopy, suggesting that there are 300 -- 500 RCB stars in the Milky Way. In our previous paper \citep{Karambelkar2021}, near-IR (NIR) \emph{J-} band lightcurves from Palomar Gattini IR \citep{De2020} together with WISE colors yielded 394 promising candidates for spectroscopic follow-up. Using NIR spectra for a 26 of these, 11 new RCB stars were presented. 

In this paper, we complete the NIR census of RCB stars and present NIR spectra (1--2.4 $\mu$m) of the full sample of RCB candidates listed in \citet{Karambelkar2021}. Additionally, we conduct a WISE-color independent search for RCB stars using PGIR lightcurves to find RCB stars missed by the color-selection criteria. We also use the PGIR lightcurves to search for new DY\,Per type stars in the Milky Way. We describe our candidate selection criteria in Section \ref{sec:cand_selection}, the NIR spectroscopic observations in Section \ref{sec:nir_spec_obs} and the classifications in Section \ref{sec:classifications}. We use this complete census of RCB stars to derive the total number of RCB stars expected in the Milky Way and infer the rate of He and CO white dwarf mergers in Section \ref{sec:discussion}. We use the NIR spectra and lightcurves to measure their radial velocities, their positions in the color-magnitude diagram and pulsation periods. We also examine high cadenced \emph{TESS} lightcurves of a sample of RCB and dLHdC stars. We conclude with a summary of our results in Section \ref{sec:summary}

\section{Candidate selection}
\label{sec:cand_selection}
\subsection{IR color + PGIR lightcurve - based selection}
Our candidate selection is described in detail in \citet{Karambelkar2021}. Briefly, we re-prioritize the WISE color-selected catalog of 2194 RCB candidates from \citet{Tisserand2020} (T20 catalog hereafter) using PGIR light-curves. First, we exclude 304 candidates that show large-amplitude periodic variations resembling AGB stars and 23 candidates whose lightcurves resemble those of RV-Tauri stars. Of the remaining candidates, we identify 177 candidates that show significant, non-periodic variations in their lightcurves, 230 candidates that show no significant variations and 253 candidates where the lightcurve is ambiguous (i.e. shows some variations that are not obviously periodic or large-amplitude). Using the location of known RCBs and LPVs in the WISE and 2MASS color diagrams, we further sub-divide Priorities A, B, C and D into seven color-based sub-categories. We prioritize categories A, 1-a, 2-a and 3-a comprising 383 candidates in total for spectroscopic follow-up. 

\begin{figure*}
    \centering
    \includegraphics[width=0.9\textwidth]{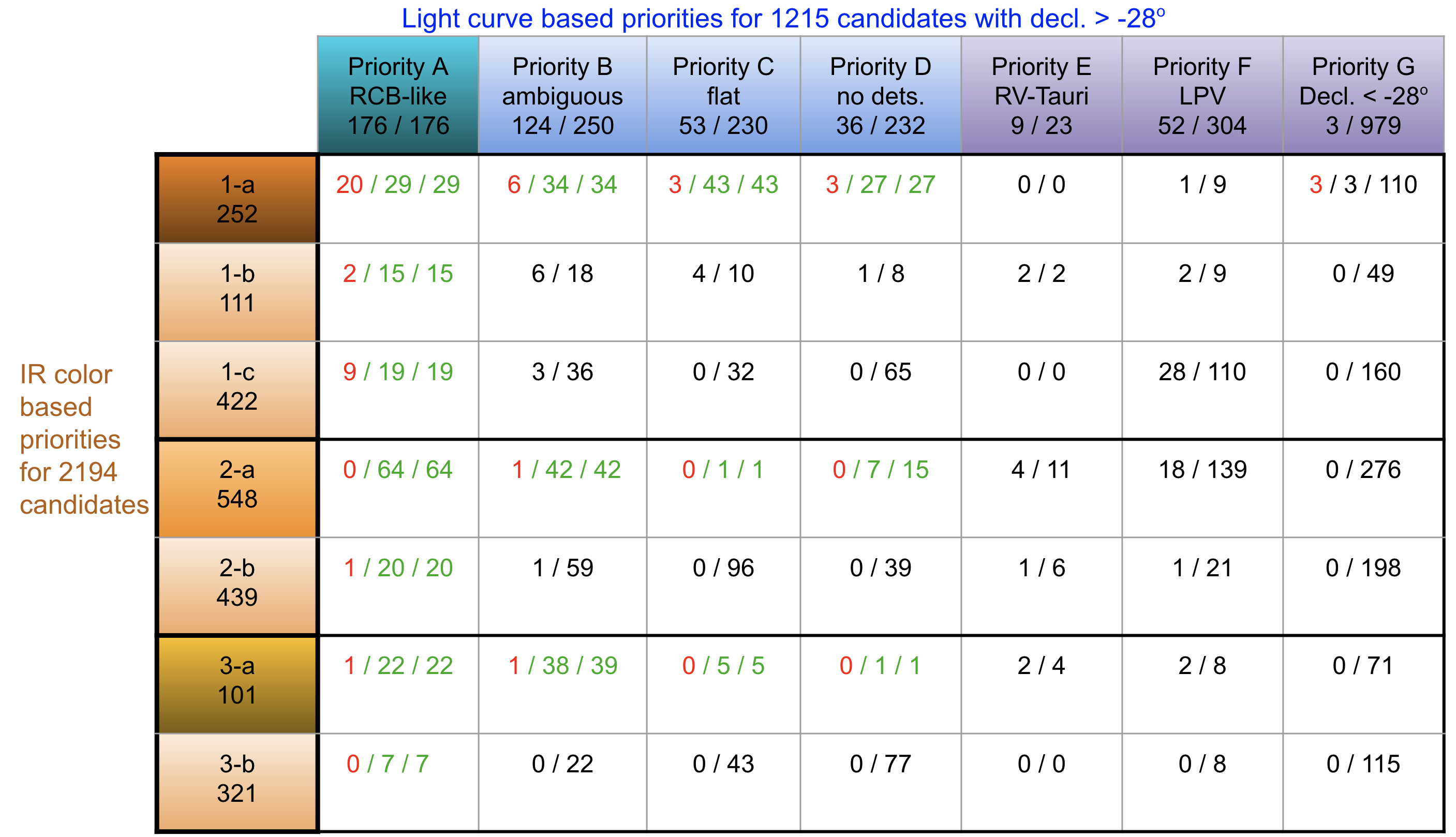}\\
    \vspace{0.3in}
    \includegraphics[width=0.5\textwidth]{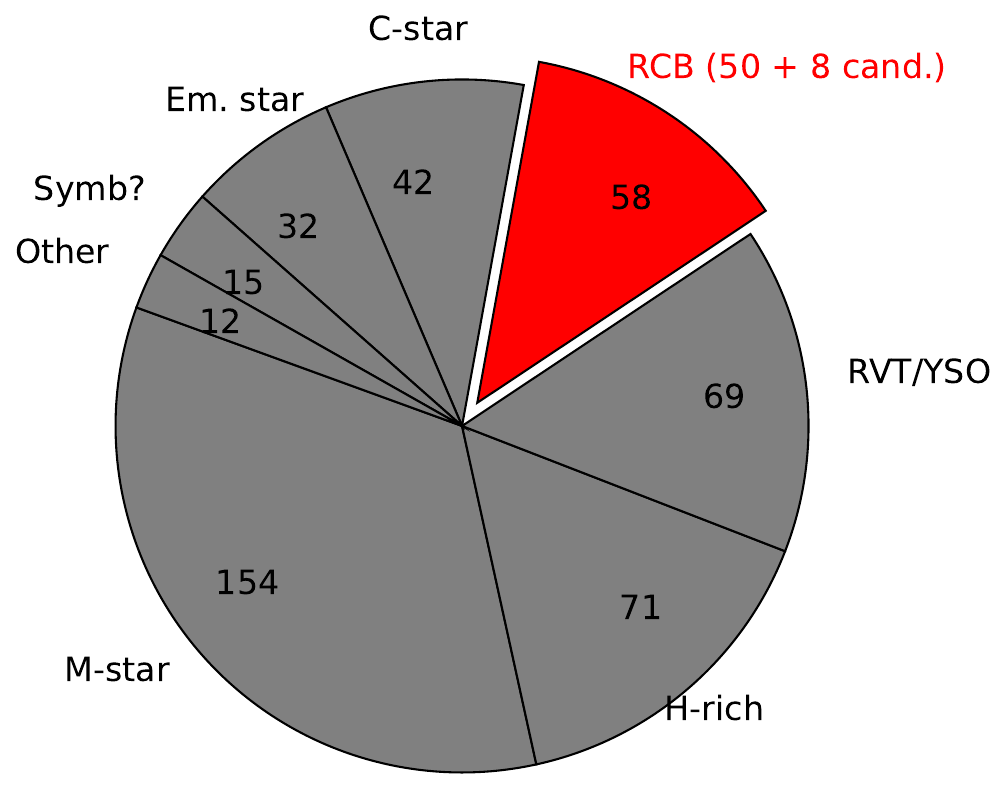}
    \caption{\emph{(Top:)} Lightcurve and color-based priorities for RCB candidates (updated from \citet{Karambelkar2021} using more recent PGIR lightcurves). The categories prioritized for spectroscopic follow-up are numbered in green. In each category, we indicate the total number of candidates and the number of candidates for which we obtained NIR spectra. We also indicate the number of RCB stars identified in each category in red. We obtained NIR spectra for a total of 453 candidates. 375 of the 383 prioritized candidates have NIR spectra, corresponding to an overall completeness of $\approx 98\%$. 3 candidates from Priority G (decl. $< -28^{o}$) were observed spectroscopically because they were listed as strong RCB candidates in T20.
    \emph{(Bottom:)} Pie-chart showing our NIR spectroscopic classifications of all 453 stars. 
    }
    \label{fig:toi_priorities_classifications}
\end{figure*}

The top panel of Fig. \ref{fig:toi_priorities_classifications} shows a table with the distribution of candidates in these categories. This table differs slightly from the one presented in \citet{Karambelkar2021}, as we updated the lightcurve-based priorities based on more recent PGIR \emph{J-}band lightcurves with additional $\approx$1.5 years of data. The categories prioritized for NIR spectroscopic followup are marked in green. We also indicate the number of candidates for which we obtained NIR spectra in each category, and the number of RCB stars spectroscopically confirmed within each category (by red numbers). The bottom panel of Fig. \ref{fig:toi_priorities_classifications} shows a pie-chart with the spectroscopic classifications of all sources we observed (discussed in Sec. \ref{sec:classifications}.) 

As the color-criteria were chosen to search for RCB stars, we do not expect to find any DY\,Per type stars in this category.

\subsection{Color independent PGIR lightcurve-based selection}
To search for possible RCB stars missed by the T20 color-selected catalog, we searched for RCB stars in the PGIR lightcurve database, independent of any color-criteria. The PGIR lightcurve database contains aperture-photometry lightcurves spanning June 2018 to July 2021 for $\sim70$ million stars that were detected in the PGIR reference images. We search for stars that show RCB-like large-amplitude, erratic variations in their PGIR lightcurves. Unfortunately, the completeness of the lightcurve database is not well-quantified, especially in dense Galactic fields that are limited by confusion noise. Therefore, we cannot use this lightcurve-selection alone for a systematic RCB search. Instead, we use this to search for possible populations of RCB stars missing from the T20 catalog.

For each lightcurve in the PGIR database, we calculate the von-neumann ratio ($\eta$, \citealp{vonNeumann1941}) and peak-to-peak amplitude (ptp). $\eta$ measures the degree of correlation between successive data-points in the lightcurve, and is well-suited to identify stars showing large-amplitude variations. Non-variable stars are expected to have $\eta$ close to 2, while smaller $\eta$ values suggest significant correlations in successive variations. We select 1200 stars with $\eta<0.5$ and ptp$>2.5$ mag (informed by known RCB stars). We then inspect their lightcurves to reject periodic variability and select 75 stars that show large-amplitude variations and are not present in the color-selected T20 catalog for further spectroscopic follow-up. As no color-information was used in the selection, we expect to find both RCB and DY\,Per type stars in this category. The spectroscopic classifications of all 75 variables will be presented in a forthcoming paper (Earley et al., in prep). Here, we focus on RCB and DY\,Per type stars found within these sources. 

\subsection{PGIR lightcurves of known carbon-stars}
As RCB stars have spectral features resembling carbon stars, we examined PGIR-J band lightcurves for 9975 stars classified as ``carbon-star" or ``candidate carbon stars" on Simbad, to search for RCB-like variations in them. We were not able to obtain NIR spectra for interesting stars identified in this category. For this reason, we report them as candidate RCB and DY\,Per type stars in Section \ref{sec:class_simbad}, pending spectroscopic confirmation.

\section{NIR Spectroscopic observations}
\label{sec:nir_spec_obs}
\begingroup
\renewcommand{\tabcolsep}{2pt}
\begin{table}
\begin{center}
\begin{minipage}{9cm}
\caption{Updated priorities of WISE-selected candidates based on PGIR lightcurves}
\label{tab:updated_priorities}
\begin{tabular}{ccccccc}
\hline
\hline
{ToI-ID} & {RA} & {Dec} & {Priority} & {} & {} &  {} \\
{}       & {deg} & {deg} & {} & {} & {} &  {} \\
\hline
23 & 37.5679 & 12.28989 & A \\
.. & & & \\
\hline
\hline
\end{tabular}
\end{minipage}
\end{center}
\end{table}
\endgroup

\begingroup
\renewcommand{\tabcolsep}{2pt}
\begin{table}
\begin{center}
\begin{minipage}{8cm}
\caption{Spectroscopic classifications of candidates selected for followup}
\label{tab:classification_catalog}
\begin{tabular}{ccccccc}
\hline
\hline
{ToI-ID} & {RA} & {Dec} & {Class.} & {Date} & {Inst.} &  {} \\
{}       & {deg} & {deg} & {} & {} & {} &  {} \\
\hline
23 & 37.5679 & 12.28989 & RV-Tauri \\
.. & & & \\
\hline
\hline
\end{tabular}
\end{minipage}
\end{center}
\end{table}
\endgroup
We obtained medium resolution NIR spectra for a total of 453 of our color-selected candidates from the T20 catalog. This includes 375 of the 383 prioritized candidates described in Section 2.1, and 81 stars from the other priorities. All but 3 observed candidates have declination $>-28^{o}$. These 3 candidates belong to Priority G and were observed because they were listed as strong RCB candidates in T20. Fig. \ref{fig:toi_priorities_classifications} shows the priority-wise distribution of stars for which we obtained NIR spectra. 


The NIR spectra were obtained on several nights from October 2019 to December 2021 with the Triplespec spectrograph (R$\approx2700$, \citealt{Herter2008}) on the 200-inch Hale telescope at Palomar Observatory and the SpeX spectrograph (R$\approx$1500) on the NASA Infrared Telescope Facility (IRTF, \citealt{Rayner2003}). The IRTF spectra were observed as part of programs 2020A111 and 2021B074. We obtained a total of 389 spectra with P200/Triplespec and 69 spectra with IRTF/SPeX All spectra were extracted using the IDL package \texttt{spextool} \citep{Cushing2004}. The extracted spectra were flux calibrated and corrected for telluric absorption with standard star observations using \texttt{xtellcor} \citep{Vacca2003}.

\section{NIR spectroscopic classifications}
\label{sec:classifications}
We classify the 453 sources using their spectra and lightcurves. We find that these sources include a mix of RCB stars, Mira variables, carbon-rich (C-rich) AGB stars, RV Tauri stars, possible giant stars in symbiotic binaries, possible young stellar objects (YSOs) and Wolf-Rayet (WR) stars. The bottom panel of Fig. \ref{fig:toi_priorities_classifications} shows the distribution of the source classifications. We discuss the classifications and properties of these sources in Appendix \ref{appendix:classifications}. Here, we focus on the sources classified as RCB stars.

As discussed in \citet{Karambelkar2021}, RCB stars are characterised by the following NIR spectral features\,--

\begin{itemize}[leftmargin=*]
    \item RCB stars show \ion{He}{1} ($\lambda 10830$) emission or absorption. The RCB stars undergoing a photometric minimum usually show \ion{He}{1} emission, however this can be suppressed by the circumstellar dust. The stars at or rising to maximum light show either a P-cygni profile or strong blueshifted absorption. 
    \item RCB stars in a minimum exhibit a mostly featureless, reddened spectrum with emission lines of He I, and sometimes Si I and C$_{2}$. 
    \item At maximum light, the spectra of RCB stars resemble F-G type supergiants, with the absence or significantly weak hydrogen lines. Prominent features include absorption lines of C I (most prominently at 1.0686 and 1.0688\um), Fe I, Si I and K I. Stars with cold effective (T$_{\rm{eff}} \leq 6800$\,K) temperatures show molecular absorption features due to CN (1.0875, 1.0929, 1.0966 and 1.0999\um), $^{12}$C$^{16}$O (2.2935, 2.3227, 2.3525, 2.3829, 2.4141\um) and $^{12}$C$^{18}$O. 
\end{itemize}


\subsection{IR color + PGIR lightcurve - based selection}
\label{sec:t20_classifications}
We examine our spectra and identify 74 stars that show some or all of the spectroscopic features described above. In cases where the spectra are not sufficient to determine a robust classification, we examine the lightcurves. We mainly use the PGIR \emph{J}-band light-curves, and also examine publicly available optical light-curves from the Zwicky Transient Facility (ZTF, \citealt{Bellm2019}) and ATLAS \citep{Tonry2018, Smith2020atlas} surveys to search for large amplitude variations. RCB stars can also show large amplitude variations at mid-IR wavelengths due to dust formation episodes. To search for these, we use publicly available 3.6 and 4.5 \um data from the NEOWISE survey. 

First, we classified 11 of these stars as RCBs in \citet{Karambelkar2021}. Of the remaining stars, 27 show RCB-like variations in the PGIR \emph{J-}band data. We classify 26 of these as RCB stars, as the NIR spectral features are consistent with the photometric phases as suggested by the light curve. The last one -- WISE-ToI-1007 shows large amplitude RCB-like variations, but does not show \ion{He}{1} in its spectrum. As the spectrum otherwise resembles that of a carbon-star, and we suggest that this is a dust-forming carbon star and not an RCB. 

26 stars do not show any large amplitude variations in PGIR data, but have spectra with features seen in those of RCB stars. Of these, we classify 8 stars as RCB stars because their spectra very closely resemble RCB stars that haven't undergone photometric declines in a long time, or the ZTF, ATLAS or NEOWISE lightcurves show clear evidence of declines or variability in the past. Of the remaining 18  sources, the spectra of 8 show strong He emission -- unlike RCB stars that haven't undergone a large-amplitude photometric decline in a long time. The ZTF and ATLAS light-curves of these 8 stars show pulsations on timescales of $\sim 10 - 100$ days superposed on smooth longer-timescale variations, similar to those seen in RV-Tauri stars \footnote{\url{https://ogle.astrouw.edu.pl/atlas/RV_Tau.html}}.These 8 stars are most likely RV-Tauri stars. Of these, we highlight the source WISE-ToI-3012, as it shows a slow rise in the optical and NIR wavelengths for the last 2000 days, and shows a large amplitude dip in the NEOWISE data. The 10 remaining stars have spectra with low S/N and no conclusive features in their lightcurves to enable a confident classification. 5 of these stars show strong He emission and some strong narrow H emission lines in their spectra, suggesting that they are likely RV-Tauri stars. We list the remaining 5 -- WISE-ToI-28, 41, 228, 293 and 1257 as candidate RCB stars. WISE-ToI-28 shows uncharacteristically large helium emission, but also shows periodic 1 mag variations in NEOWISE. WISE-ToI-293 shows a short timescale decline in ZTF data. WISE-ToI-41 shows weak He emission and erratic NEOWISE variations.  WISE-ToI-1257 shows a steep rising red continuum with small variations in the \emph{J}-band.

7 stars have no detections in the PGIR data: WISE-ToI-195, 270, 245, 288, 317, 321 and 323. Their spectra show the helium emission line. No strong hydrogen features are evident, but we cannot rule out their presence as the spectra have low S/N. Of the seven, WISE-ToI-195 and WISE-ToI-270 show RCB-like declines in their ZTF lightcurves. We classify these two as RCB stars. WISE-ToI-317 and 321 do not show any significant declines in their ZTF and ATLAS lightcurves. These two stars have distance estimates from \emph{Gaia} \citep{Bailer-Jones2018} which suggest that their absolute magnitudes are $M_{r}\approx$ 4.1 and 1.8 respectively -- inconsistent with them being RCB stars. For this reason, we exclude them as RCB stars\footnote{We note that the significance of the \emph{Gaia} parallax measurement for WISE-ToI-317 is low (Plx/e\_Plx = 2.9). Nevertheless, there is no compelling evidence, so we do not classify this as an RCB star.}. The remaining three -- WISE-ToI-245, 288 and 323 do not show any declines in their ZTF and ATLAS lightcurves, and do not have any additional information from Gaia. We list these three as strong RCB candidates. 


Finally, we observed 3 stars -- WISE-ToI-164, 181 and 264 -- that lie outside the area covered by PGIR (i.e. have $\delta<-28^{\rm{o}}$ and belong to Priority G), but are listed as strong RCB candidates in \citet{Tisserand2020}. Our NIR spectra confirm that they are RCB stars. 

\begin{figure*}
    \centering
    \includegraphics[width=\textwidth]{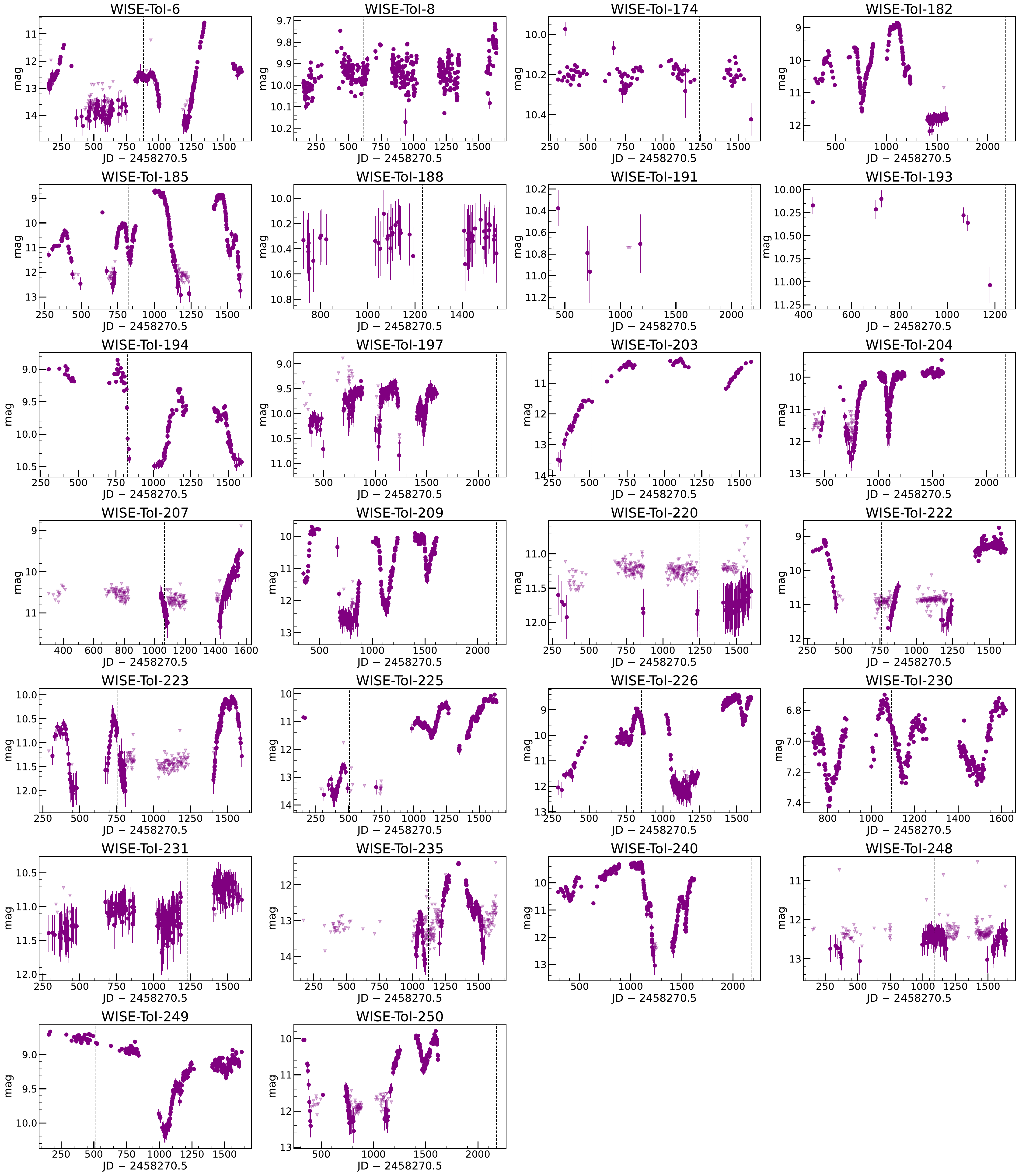}
    \caption{PGIR \emph{J-}band lightcurves of new RCB stars. Circles denote detections, while triangles denote upper limits (continued in Fig. \ref{fig:newrcbs_2_lcs}). The dotted black vertical line marks the epoch when the NIR spectrum was obtained.}
    \label{fig:newrcbs_1_lcs}
\end{figure*}

\begin{figure*}
    \centering
    \includegraphics[width=\textwidth]{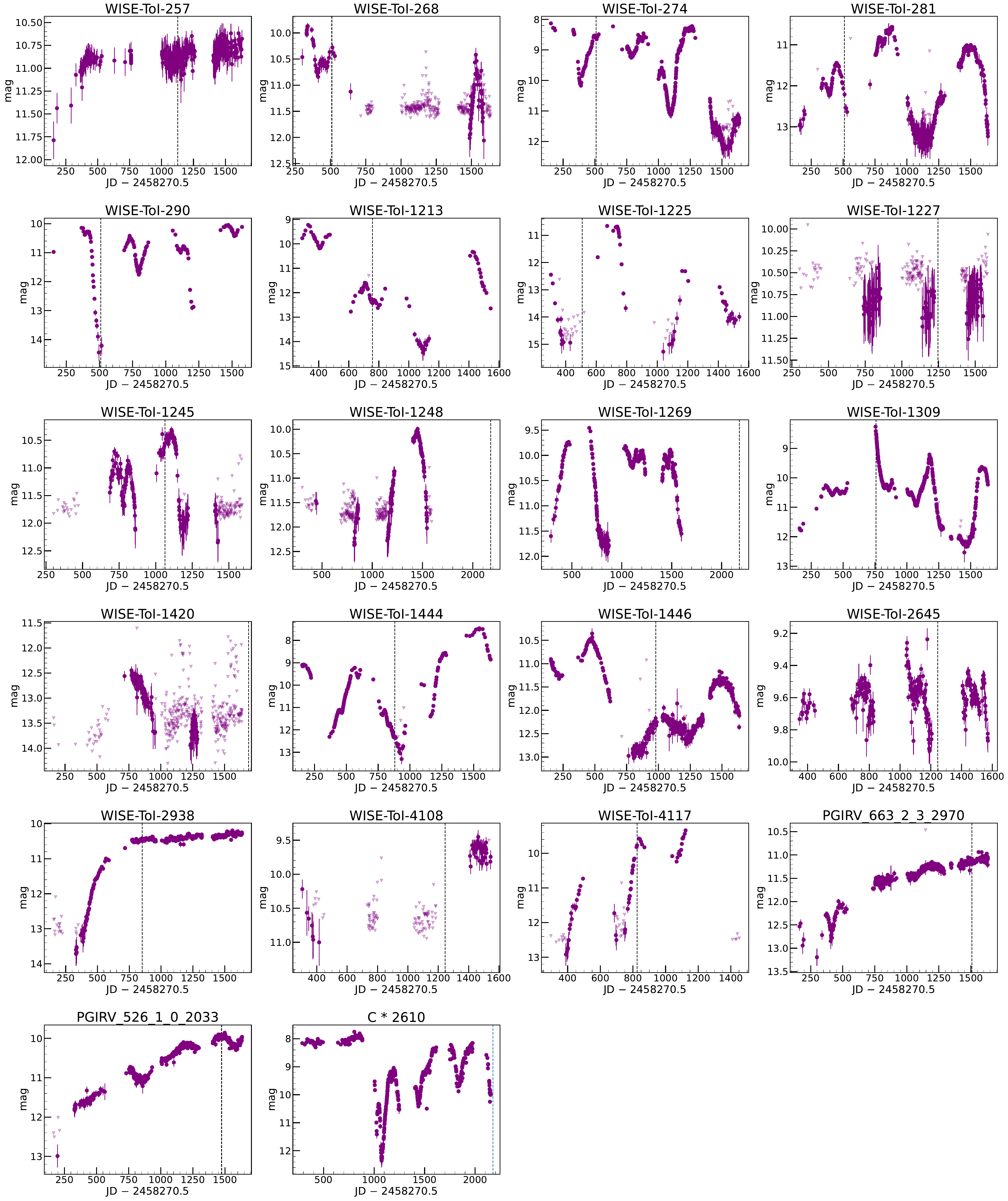}
    \caption{PGIR \emph{J-}band lightcurves of new RCB stars}
    \label{fig:newrcbs_2_lcs}
\end{figure*}

\begin{figure*}
    \centering
    \includegraphics[width=\textwidth]{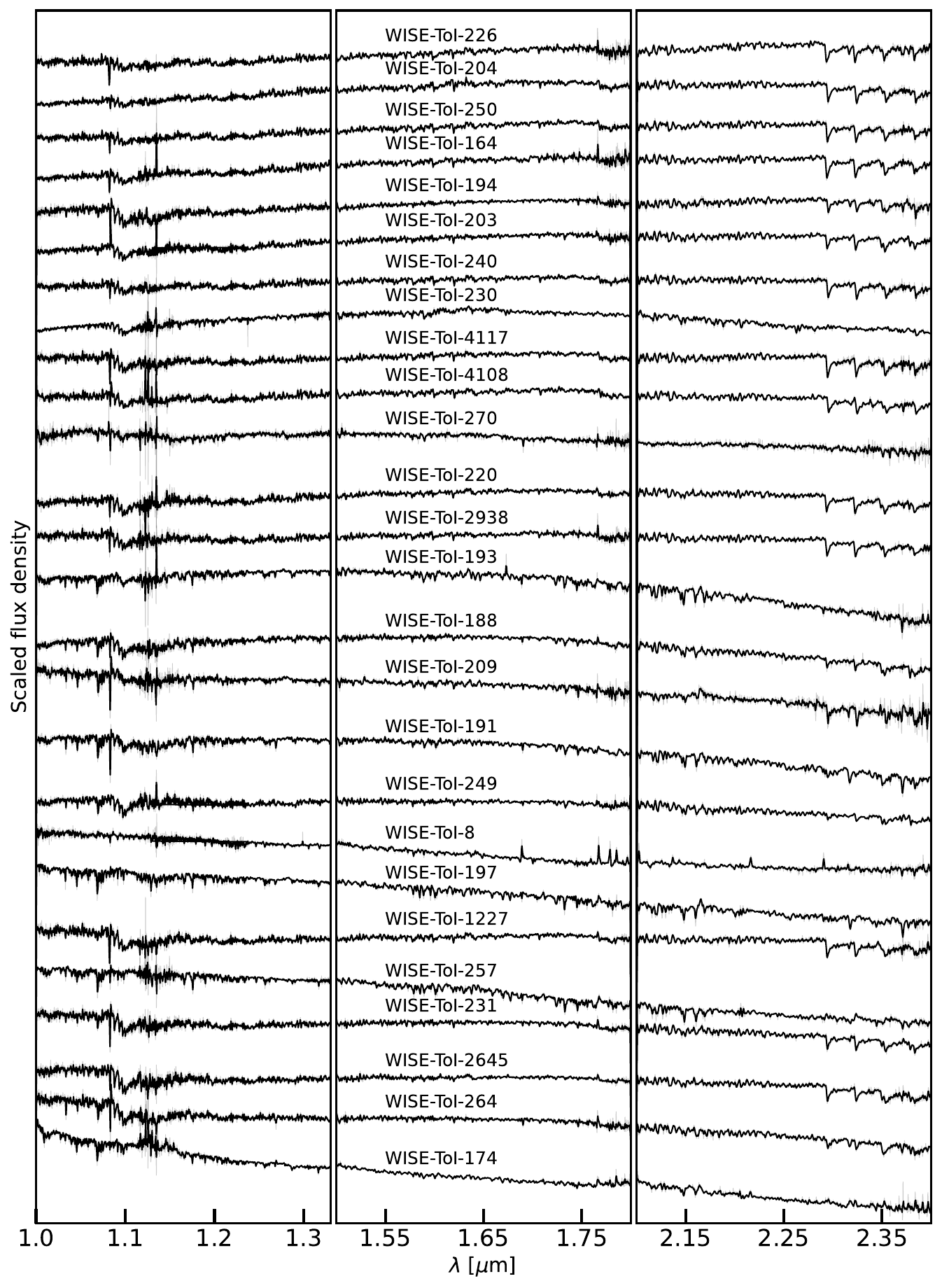}
    \caption{NIR spectra of new RCB stars}
    \label{fig:newrcbs_1_spectra}
\end{figure*}

\begin{figure*}
    \centering
    \includegraphics[width=\textwidth]{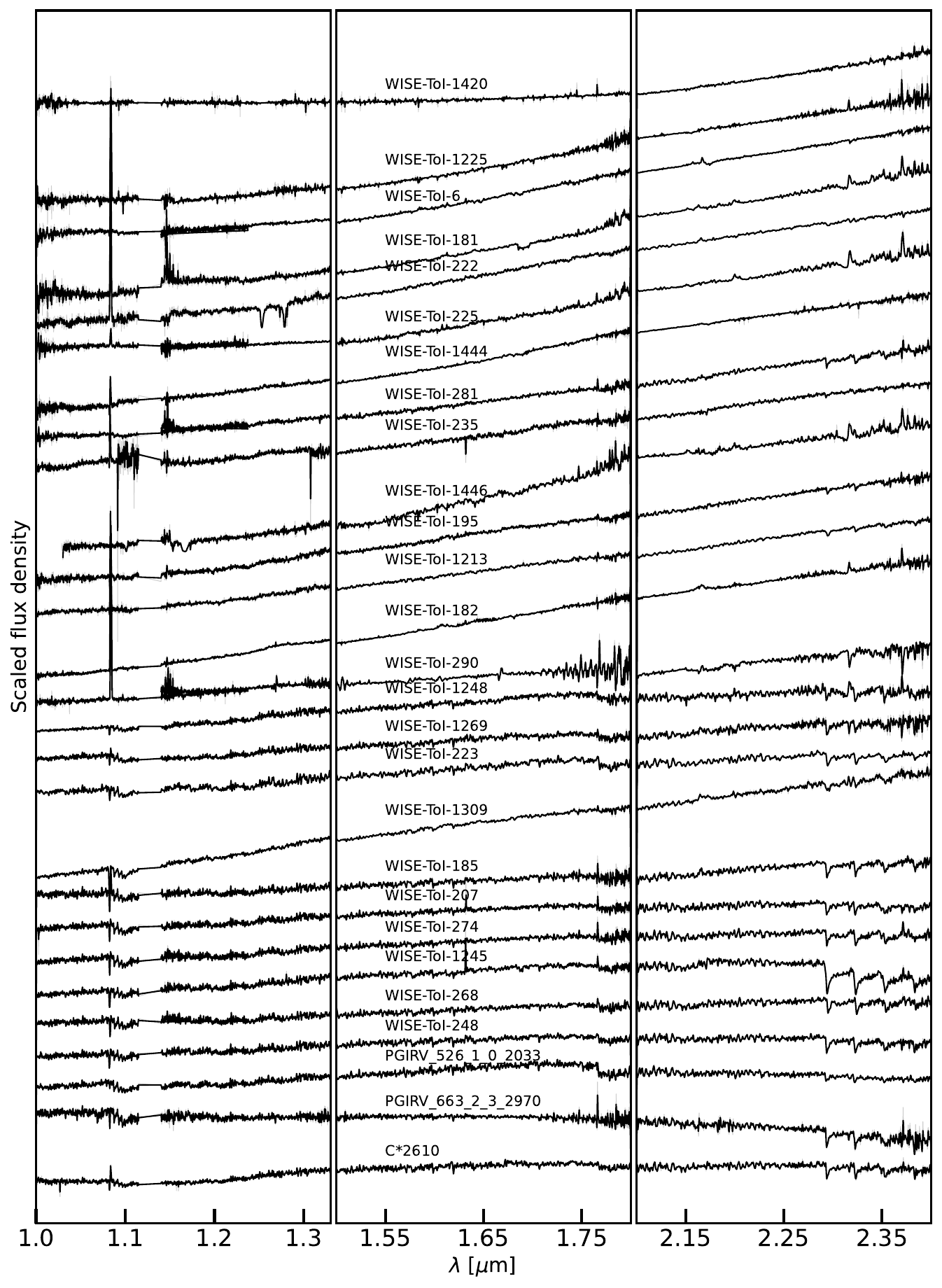}
    \caption{NIR spectra of new RCB stars (continued from Fig. \ref{fig:newrcbs_1_spectra})}
    \label{fig:newrcbs_2_spectra}
\end{figure*}

\begin{figure*}
    \centering
    \includegraphics[width=\textwidth]{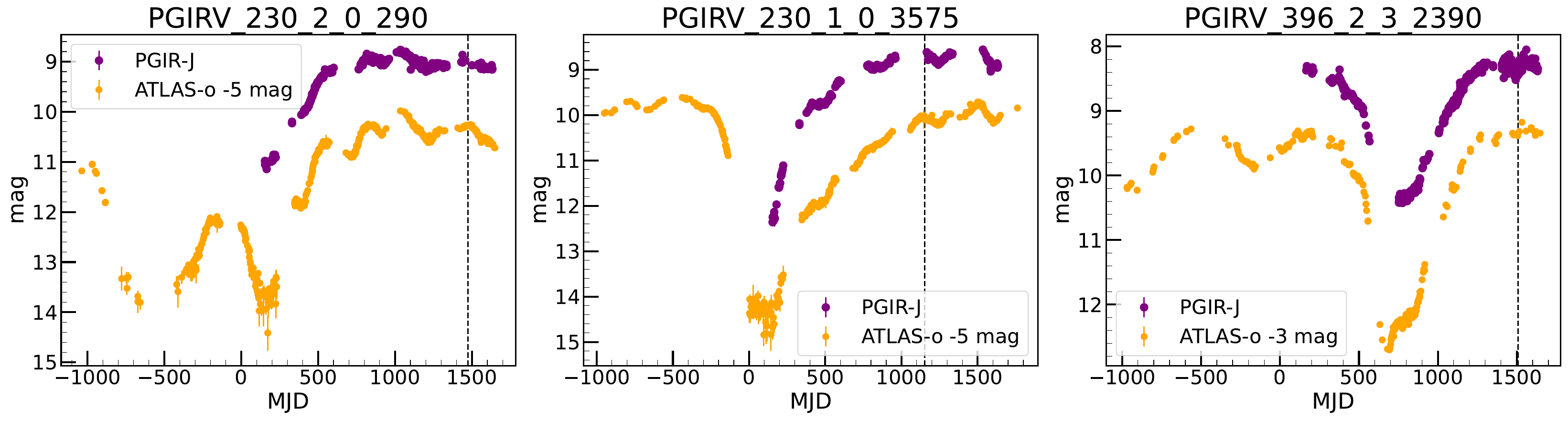}
    \includegraphics[width=\textwidth]{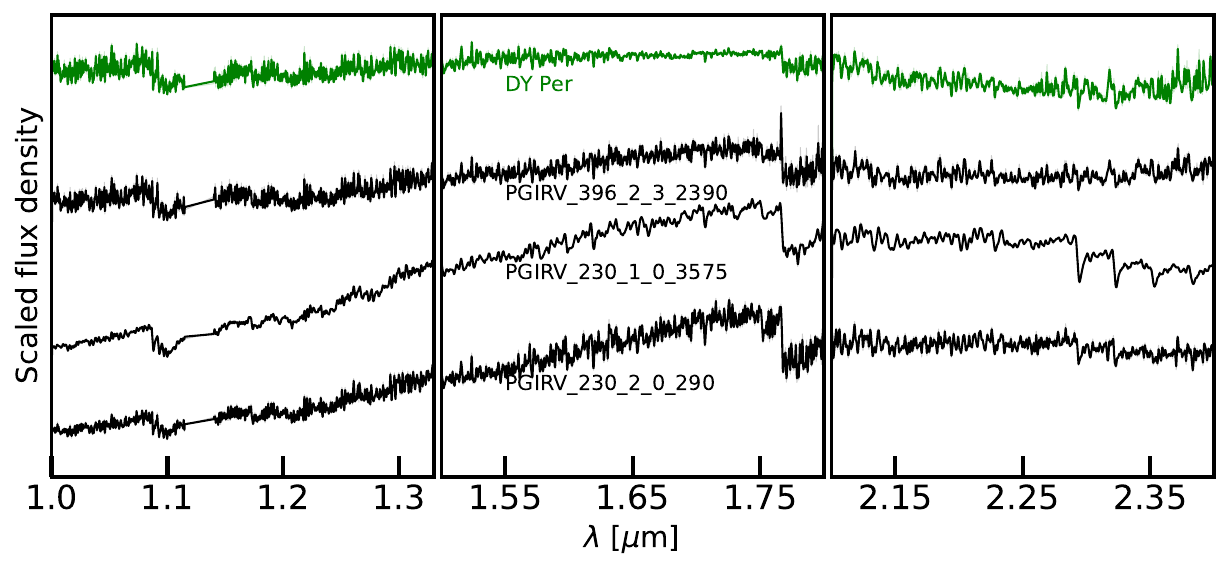}
    \caption{\emph{Top:} Lightcurves of new DYPer type stars. Purple dots show the PGIR data and orange dots show the shifted ATLAS-o band data. \emph{Bottom:} NIR spectra of the new DYPer type stars (black) and DY\,Per itself (green). The spectra of the new stars closely resemble that of DY\,Per and N-type carbon stars.}
    \label{fig:dyper_lcs}
\end{figure*}

To summarize, we identify a total of 50 RCB stars and 8 strong RCB candidates using the PGIR-lightcurve-based prioritization of the T20 catalog. We note that of the 50 RCB stars, 19 were identified previously as RCB stars from their optical spectra by \citet{Tisserand2020}. Our NIR spectra independently confirm their nature as RCBs. 6 additional RCB stars were listed as strong candidates by \citet{Tisserand2020}, which we unambiguously classify as RCB stars based on their NIR spectra. Two stars WISE-ToI-185 and WISE-ToI-226 were listed as a potential RCBs based on their optical lightcurves by \citet{Eyer2023} and \citet{Apellaniz2023} respectively, and are now spectroscopically classified as RCBs based on our NIR spectra. All 50 RCB stars identified or confirmed by our NIR spectra are listed in Table \ref{tab:rcbs_table} and \ref{tab:rcbs_table_2}. The PGIR lightcurves of the 45 RCB stars that have PGIR detections are plotted in Figures \ref{fig:newrcbs_1_lcs}, \ref{fig:newrcbs_2_lcs}. The NIR spectra of all 50 RCB stars are plotted in Figures \ref{fig:newrcbs_1_spectra} and \ref{fig:newrcbs_2_spectra}. The 8 strong RCB candidates are listed in Table \ref{tab:rcbs_cands_table}, and their lightcurves and spectra are shown in Fig. \ref{fig:new_cand_rcbs_2_spectra}.

\subsection{Color-independent, PGIR lightcurve-only selection}
\subsubsection{New RCB stars}
Of the 75 lightcurve-selected candidates, we identify 2 new RCB stars PGIRV663-2-3-2970 (PGIRV663 hereafter) and PGIRV526-1-0-2033 (PGIRV526 hereafter). Their lightcurves and spectra are included in Fig. \ref{fig:newrcbs_2_lcs} and \ref{fig:newrcbs_2_spectra} respectively. The \emph{J-}band lightcurves of both these stars shows a slow rise of $\approx$3 mag over 1500 days. The ZTF lightcurve of PGIRV663 shows a 2 mag. decline, 1700 days prior to the epoch of our NIR spectrum. Based on the ZTF lightcurve, this candidate was flagged as a possible RCB star on AAVSO by G. Murakawski. The NIR spectra of both stars are similar to RCB stars that are recovering from a photometric decline, consistent with their lightcurves. Both stars have WISE and 2MASS detections, but were missed by the color-selection criteria of \citet{Tisserand2020}. PGIRV526 has $W2-W3 = 0.85$ -- failing the criterion $W2-W3 >1.1$ mag, suggesting that it has hot dust (T$>1000$ K) around it. PGIRV663 has $J-H = 0.70$ and $H-K = 0.43$ mag, failing the criterion $J-H < (H-K) + 0.2$  -- suggesting that it is a warm RCB star with a thin dust shell. 

\begingroup
\renewcommand{\tabcolsep}{10pt}
\begin{table*}
\begin{center}
\begin{minipage}{16cm}
\caption{RCB and DY\,Per stars identified/confirmed from our NIR census (continued in Table \ref{tab:rcbs_table_2})}
\label{tab:rcbs_table}
\begin{tabular}{lccccc}
\hline
\hline
{Name} & {ToI--ID/} & {RA} & {Dec} & {NIR spec.} & {Comments} \\
{}     &  {PGIR Name}      & {deg} & {deg} & {class} & {}  \\
\hline
WISE J174317.52-182402.4 &  185 & 265.82303 & -18.40068 & RCB & a\\  
WISE J175317.73-194632.5 &  195 & 268.32389 & -19.77572 & RCB & \\ 
WISE J182501.85-230803.9 &  226 & 276.25772 & -23.13444 & RCB & b\\ 
WISE J182801.05-100916.7 &  230 & 277.00438 & -10.15464 & RCB & \\  
WISE J183213.53+050454.5 &  235 & 278.05638 & 5.08181 & RCB & \\ 
WISE J184102.48-004136.3 &  248 & 280.26034 & -0.69344 & RCB & \\ 
WISE J190918.81+030531.2 &  270 & 287.32838 & 3.09201 & RCB & \\  
WISE J175725.03-230426.4 & 1245 & 269.35429 & -23.07402 & RCB & \\ 
WISE J180021.11-232202.9 & 1248 & 270.08797 & -23.36749 & RCB &  \\
WISE J203825.90+514140.7 & 1420 & 309.60795 & 51.69464 & RCB & \\ 
WISE J221558.89+422246.2 & 1444 & 333.99538 & 42.37950 & RCB & \\ 
WISE J223517.61+593812.7 & 1446 & 338.82341 & 59.63688 & RCB & \\  
WISE J202514.28+472731.5 & 2938 & 306.30950 & 47.45877 & RCB & \\ 
WISE J181706.84-235751.3 & 4108 & 274.27851 & -23.96426 & RCB & h\\  
WISE J175136.80-220630.6 &  194 & 267.90335 & -22.10852 & RCB & c\\
WISE J180313.12-251330.1 &  207 & 270.80467 & -25.22505 & RCB & c\\ 
WISE J183631.25-205915.1 & 4117 & 279.13024 & -20.98755 & RCB & c \\
WISE J172044.89-315031.7 &  164 & 260.18707 & -31.84215 & RCB & c, g \\
WISE J173837.00-281734.5 &  181 & 264.65417 & -28.29292 & RCB & c, g \\
WISE J190309.89-302037.1 &  264 & 285.79123 & -30.34365 & RCB & c, g \\
WISE J004822.34+741757.4 &    6 &  12.09309 & 74.29928 & RCB & e\\ 
WISE J005128.08+645651.7 &    8 &  12.86702 & 64.94770 & RCB & e\\ 
WISE J175749.76-075314.9 &  203 & 269.45737 & -7.88750 & RCB & d, e\\ 
WISE J181836.38-181732.8 &  222 & 274.65160 & -18.29247 & RCB & e\\ 
WISE J182010.96-193453.4 &  223 & 275.04570 & -19.58150 & RCB & e\\ 
WISE J182235.25-033213.2 &  225 & 275.64690 & -3.53701 & RCB & e\\ 
WISE J184158.40-054819.2 &  249 & 280.49336 & -5.80535 & RCB & d, e\\ 
WISE J190813.12+042154.1 &  268 & 287.05469 & 4.36503 & RCB & e\\ 
WISE J191243.06+055313.1 &  274 & 288.17945 & 5.88698 & RCB & e\\
WISE J192348.98+161433.7 &  281 & 290.95410 & 16.24270 & RCB & e\\ 
WISE J194218.38-203247.5 &  290 & 295.57660 & -20.54654 & RCB & d, e\\ 
WISE J170552.81-163416.5 & 1213 & 256.47005 & -16.57125 & RCB & e\\ 
WISE J173737.07-072828.1 & 1225 & 264.40446 & -7.47449 & RCB & e\\ 
WISE J185726.40+134909.4 & 1309 & 284.36004 & 13.81930 & RCB & e\\ 
WISE J172951.80-101715.9 &  174 & 262.46586 & -10.28778 & RCB & d \\
WISE J174645.90-250314.1 &  188 & 266.69128 & -25.05392 & RCB & d \\
WISE J175107.12-242357.3 &  193 & 267.77967 & -24.39927 & RCB & d \\
WISE J181538.25-203845.7 &  220 & 273.90938 & -20.64604 & RCB & d \\
WISE J182943.83-190246.2 &  231 & 277.43263 & -19.04617 & RCB & d \\
WISE J185525.52-025145.7 &  257 & 283.85636 & -2.86271  & RCB & d \\
WISE J173819.81-203632.1 & 1227 & 264.58255 & -20.60893 & RCB & d\\ 
WISE J181252.50-233304.4 & 2645 & 273.21875 & -23.55124 & RCB & d\\
WISE J182723.38-200830.1 & 1269 & 276.84744 & -20.14172 & RCB & d\\ 
WISE J180550.49-151301.7 &  209 & 271.46038 & -15.21714 & RCB & d\\ 
WISE J175558.51-164744.3 &  197 & 268.99382 & -16.79565 & RCB & d\\  
WISE J175749.98-182522.8 &  204 & 269.45827 & -18.42300 & RCB & d\\ 
WISE J175031.70-233945.7 &  191 & 267.63210 & -23.66270 & RCB & d\\ 


\hline
\hline
\end{tabular}
\begin{tablenotes} 
\item $a$ : Listed as a RCB candidate based on the \emph{Gaia} lightcurve by \citet{Eyer2023}. $b$ : Listed as a RCB candidate based on its lightcurve by \citet{Apellaniz2023}. $c$ : Listed as strong RCB candidates by \citet{Tisserand2020}, $d$: Classified as RCB stars from optical spectra by \citet{Tisserand2020}, $e$: Also presented in our previous pilot NIR spectroscopic paper \citet{Karambelkar2021}, $f$: Flagged as a possible RCB star on AAVSO by Gabriel Murakawski. $g$: Targets with declination $<-28^{o}$ that were observed as they were listed as strong RCB candidates, \textbf{$h$: Listed as RCB star V2331 Sgr in \citet{Crawford2023}}.
\end{tablenotes}

\end{minipage}
\end{center}
\end{table*}
\endgroup

\begingroup
\renewcommand{\tabcolsep}{10pt}
\begin{table*}
\begin{center}
\begin{minipage}{18cm}
\caption{RCB and DY\,Per stars identified/confirmed from our NIR census (continued from Table \ref{tab:rcbs_table})}
\label{tab:rcbs_table_2}
\begin{tabular}{lccccc}
\hline
\hline
{Name} & {ToI--ID/} & {RA} & {Dec} & {NIR spec.} & {Comments} \\
{}     &  {PGIR Name}      & {deg} & {deg} & {class} & {}  \\
\hline
WISE J184246.26-125414.7 &  250 & 280.69277 & -12.90409 & RCB & d\\ 
WISE J174138.87-161546.4 &  182 & 265.41199 & -16.26291 & RCB & d\\ 
WISE J183649.54-113420.7 &  240 & 279.20645 & -11.57244 & RCB & d\\ 
\hline
IRAS 19437+2812          & PGIRV\_526\_1\_0\_2033 & 296.4334005	& 28.33476509 & RCB \\
MGAB-V209                & PGIRV\_663\_2\_3\_2970 & 288.3546954 & 17.61718824 & RCB & f\\
C\,*\,2610 & & 278.80827 & -15.60382 & RCB & a \\
\hline
IRAS 21210+4922          & PGIRV\_230\_2\_0\_290  & 320.699761  & 49.58775202 & DY\,Per & \\
BC 279                   & PGIRV\_230\_1\_0\_3575 & 316.0404598	& 51.96017409 & DY\,Per & \\
NC50\_6                 & PGIRV\_396\_2\_3\_2390 & 300.5364042	& 36.46534761 & DY\,Per & \\

 
\hline
\hline
\end{tabular}
\begin{tablenotes} 
\item $a$ : Listed as a RCB candidate based on the \emph{Gaia} lightcurve by \citet{Eyer2023}. $b$ : Listed as a RCB candidate based on its lightcurve by \citet{Apellaniz2023}. $c$ : Listed as strong RCB candidates by \citet{Tisserand2020}, $d$: Classified as RCB stars from optical spectra by \citet{Tisserand2020}, $e$: Also presented in our previous pilot NIR spectroscopic paper \citet{Karambelkar2021}, $f$: Flagged as a possible RCB star on AAVSO by Gabriel Murakawski. 
\end{tablenotes}

\end{minipage}
\end{center}
\end{table*}
\endgroup

\subsubsection{New DY\,Per type stars}
DY\,Per type stars are thought to be a colder sub-class of RCB stars (with T$_{\rm{eff}} \approx 3500$\,K), marked by shallower and more symmetric declines in their lightcurves than RCB stars \citep{Alcock2001}. From our PGIR lightcurve-selected candidates, we identified three stars that show lightcurves and spectra resembling DY\,Per type stars.  Fig. \ref{fig:dyper_lcs} show the PGIR-J band and ATLAS-o band lightcurves of these stars. For two of them, the PGIR lightcurves sample only the rise out of the decline, but the longer baseline ATLAS lightcurves show DY\,Per-like variations. The NIR spectra of these stars very closely resemble N-type carbon star templates from the IRTF spectral library \citep{Rayner2003}. All three stars have NIR colors similar to known DY\,Per type stars \citep{Tisserand2013}. We classify these three as new Galactic DY\,Per type stars. 

\subsection{PGIR lightcurves of known carbon-stars}
\label{sec:class_simbad}
\subsubsection{New RCB star}
Using the PGIR lightcurves of stars classified as ``carbon-stars" on Simbad, we identified one previously unknown RCB candidate -- C*2610 -- that shows large-amplitude RCB-like declines in its lightcurve. We classify it as an RCB star based on its NIR spectrum which resembles RCB stars undergoing declines. This source was also listed as a possible RCB star based on its \emph{Gaia} lightcurve by \citet{Eyer2023}, and appears in the carbon-star catalogs of \citet{Stephenson1973} and \citet{Alksnis2001}. This star is listed in Table \ref{tab:rcbs_table} and its PGIR lightcurve and NIR spectra are included in Fig. \ref{fig:newrcbs_2_lcs} and \ref{fig:newrcbs_2_spectra} respectively. 

\subsubsection{New DY\,Per candidates}
From the Simbad ``carbon-star" lightcurves, we identify 15 stars that show variations resembling DY\,Per. We list these stars as candidate DY\,Per type stars in Table \ref{tab:rcbs_cands_table}, as we do not have spectroscopic observations for them. These PGIR lightcurves of these stars are shown in Fig. \ref{fig:dyper_cands_lcs}.

\begin{figure*}
    \centering
    \includegraphics[width=0.5\textwidth]{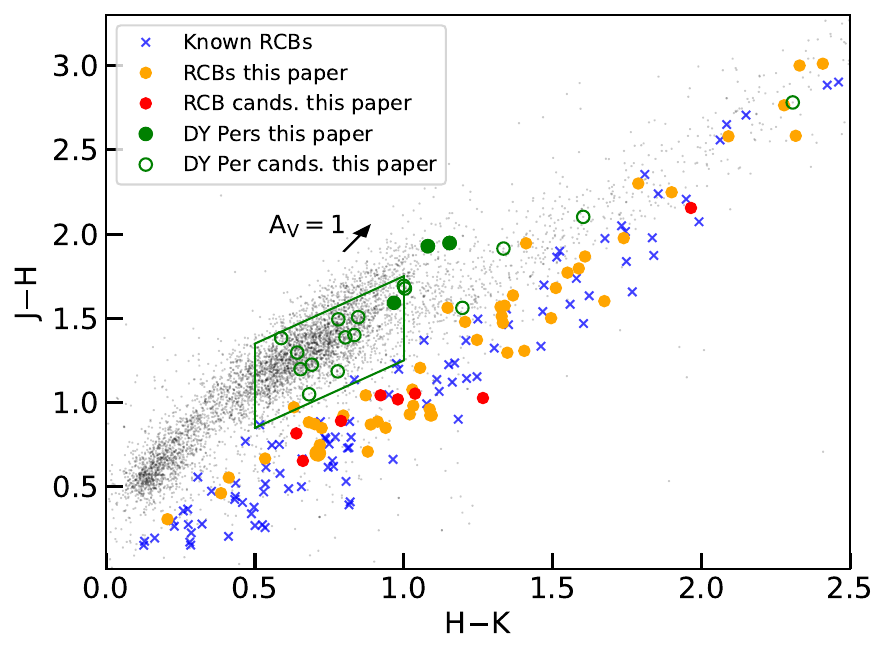}\\
    \includegraphics[width=0.9\textwidth]{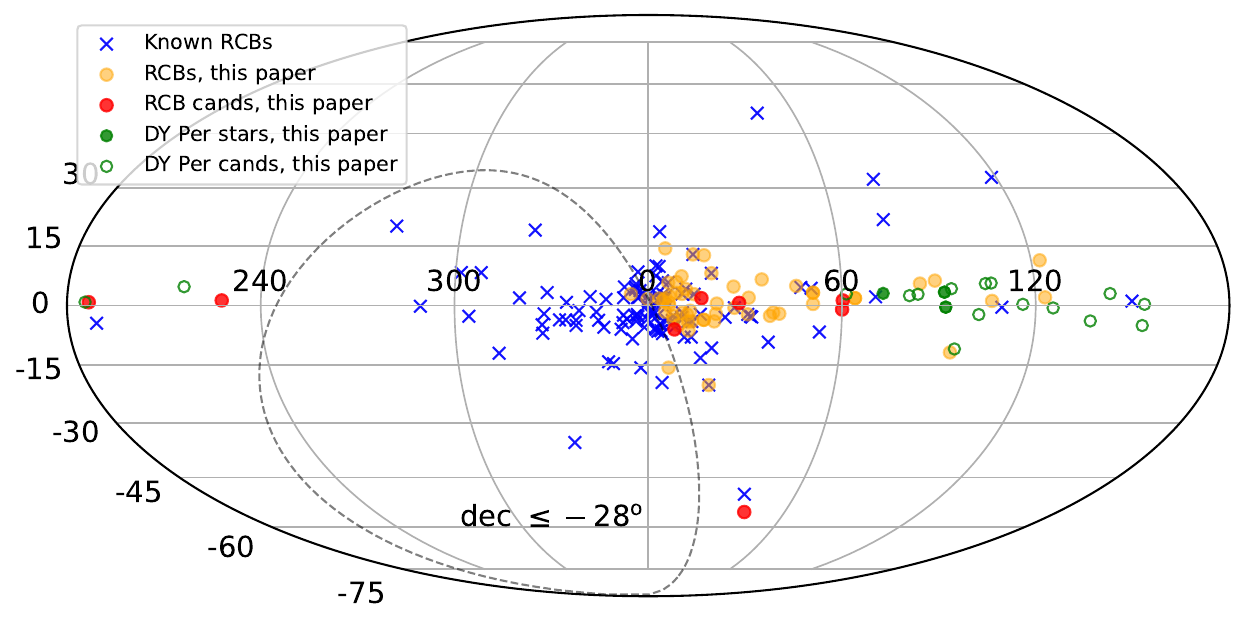}
    \caption{\emph{Top:} 2MASS color-color diagram for known RCB stars (blue crosses), new RCB stars (orange dots), new RCB candidates (red dots), new DY\,Per type stars (green solid dots) and new DY\,Per candidates (hollow green circles). The green box marks the DY\,Per-selection region from \citet{Tisserand2013}, and the black arrow indicates the direction of interstellar extinction of A$_{V}=1$ mag. The DY\,Per type stars follow a distinct trend than the RCB stars. \emph{Bottom :} Galactic distribution of  RCB and DY\,Per type stars. Most RCB stars are located towards the center of the Milky Way, consistent with a bulge population, while most DY\,Per type stars are located at high Galactic longitudes, indicative of a disk population.}
    \label{fig:rcb_dyper_colcol_dist}
\end{figure*}

\section{Discussion}
\label{sec:discussion}
\subsection{Total number of Galactic RCB stars}

\subsubsection{Total RCB stars in the T20 catalog}
\label{sec:num_rcbs_t20}
First, we determine the total number of RCB stars in the T20 catalog. We identified a total of 50 RCB stars (see Sec. \ref{sec:t20_classifications} and Fig. \ref{fig:toi_priorities_classifications}). As noted earlier, 3 of these belong to Priority G (decl.$<-28^{o}$) which is not amongst our prioritized categories, so we do not use these to determine the total number. The remaining 47 RCB stars come from our systematic followup of prioritized categories highlighted in Fig. \ref{fig:toi_priorities_classifications}. 

For lightcurve Priority A (see Fig. \ref{fig:toi_priorities_classifications}), we observed all 176 candidates and identified 33 RCB stars. For lightcurve priorities (lc-pri) B, C and D, we only observed the color-priorities (col-pri) 1-a, 2-a and 3-a. We use these observations to estimate the numbers in col-pri 1-b, 2-b and 3-b. 

First, in col-pri 1-a and lc-pri B+C+D, we observed all 104 candidates and identified 12 RCB stars. For col-pri 1-b and lc-pri B+C+D, we observed 11 out of 36 candidates and identified no RCB stars, but 1 strong RCB candidate. Scaling to the total number of candidates in this category, we expect no more than $\approx$3 RCB stars in 1-b. The efficiency is expected to be lower in 1-b than 1-a as 1-b is more contaminated by RV-Tauri stars. For col-pri 1-c and lc-pri B+C+D, we expect the contamination of LPVs to be much higher than 1-a+1-b. Based on the estimate in \citet{Karambelkar2021}, we expect 1-c to have 5 times more LPVs and 10 times fewer RCBs than 1-a and 1-b combined. Therefore, for the ambiguous and non-detection lightcurves (lc-pri B and D) in 1-c, we estimate the RCB-occurrence rate to be 50 times lower than 1-a and 1-b. For flat lightcurves (lc-pri C), we assume the same RCB-occurrence rate in 1-c as in 1-a + 1-b. This gives a total expected number of 3 RCB stars in 1-c.

Second, in col-pri 2-a and lc-pri B+C+D, we observed 50 candidates and identified 1 RCB star. Assuming this RCB-occurrence rate, we expect 1 RCB star in col-pri 2-a and 4 RCB stars in col-pri 2-b.  

\begingroup
\renewcommand{\tabcolsep}{6pt}
\begin{table}
\begin{center}
\begin{minipage}{8cm}
\caption{Total number of RCB stars identified in the T20 catalog for different color and lightcurve priorities. * marks categories which were not covered in our spectroscopic followup. The numbers in these categories were determined as described in Sec. \ref{sec:num_rcbs_t20}}
\label{tab:t20_rcb_numbers}
\begin{tabular}{ccc}
\hline
\hline
{col-pri} & {lc-pri} & {lc-pri} \\
{} & {A} & {B+C+D} \\
\hline
1-a & 20 & 12 \\
1-b & 2 & 3* \\
1-c & 9 & 3* \\
2-a & 0 & 1 \\
2-b & 1 & 4* \\
3-a & 1 & 1 \\
3-b & 0 & 29* \\
\hline
\hline
\end{tabular}
\end{minipage}
\end{center}
\end{table}
\endgroup

Finally, for col-pri 3-a and lc-pri B+C+D, we observed 44 out of 45 candidates and identified 1 RCB star. The low efficiency is expected, as these are the brightest targets in group 3, which coincide with stars with a classification listed on Simbad. Category 3-b groups the candidates with at least one upper limit in the 4 WISE bands. Optimistically, we choose an RCB efficiency of $\approx20\%$ for this category (following \citealt{Tisserand2020}), we estimate another 29 RCB stars from colpri 3-b and lc-pri B+C+D to account for a population of highly dust enshrouded  RCB stars. Table \ref{tab:t20_rcb_numbers} shows a summary of number of RCB stars in each of the categories discussed above.

In total, we estimate there are $\approx$86 RCB stars (95\% confidence interval 60 -- 150) out of the 1215 candidates from the T20 catalog with northern declinations ($\delta > -28 ^{\rm{o}}$). Accounting for southern candidates, 85\% completeness and adding the 77 known RCB stars gives a total of 86 $\times$  (100/85) $\times$ (2194/1215) + 77 = 260 (95 \% confidence interval 200 -- 390). Including the 8 strong RCB candidates gives a total of 280 (95\% CI 210 -- 400) RCB stars in the full T20 catalog.


\subsubsection{Total Galactic RCB stars}
We now discuss possible biases associated with our search, and correct for them to determine the total number of Galactic RCB stars. First, as noted in \citet{Tisserand2020}, the detection efficiency of the catalog drops within a few degrees of the Galactic center and along
the Galactic plane at low Galactic latitude due to high interstellar
extinction. We used the white-dwarf binary population synthesis model from \citet{Lamberts2019} to estimate the number of RCB stars within this region. From the simulated white-dwarf binaries, we calculate the number of He-CO WD binaries that are within 2 degrees of the Galactic center, or within 1$^{\rm{o}}$ of the Galactic plane. We find that $\approx20\%$ of all He-CO WD binaries lie within this region. Assuming that RCB stars follow a similar distribution, we estimate that there are $\approx 70$ RCB stars within this region that are missed by our search. We derive a similar number using star counts generated from the Besancon model of the Milky Way \citep{Czejak2014}. Future high-spatial resolution observations (e.g. with the \emph{Roman} space telescope) can accurately measure the number of RCB stars in this highly crowded region. Our rough estimate suggests a total of $\approx 350$ (C.I. 250 -- 500) RCB stars in the Milky Way. This estimate agrees well with the estimate of 300 -- 500 from \citet{Tisserand2020}. 

Second, the T20 catalog is expected to contain 85\% of RCB stars based on previously known RCB stars, that come from a non-homogenous sample. We use our color-independent PGIR lightcurve-based search to test this 85\% completeness estimate. After applying our lightcurve selection criteria ($\eta<0.5$ and ptp>2) on lightcurves in the PGIR database, we recover a total of 24 RCB stars. 22 of them are either previously known RCB stars or present in the T20 catalog. Only 2 RCBs are previously unknown and are not present in the catalog. Other than these 2, we do not find any RCBs in the lightcurve selected candidates. This suggests that the $85\%$ completeness estimate of the T20 catalog is reasonable. We note that the completeness of the PGIR lightcurve database is not well quantified, but it is unlikely that we are missing a substantial population of RCB stars. 

\begin{figure}
    \centering
    \includegraphics[width=0.45\textwidth]{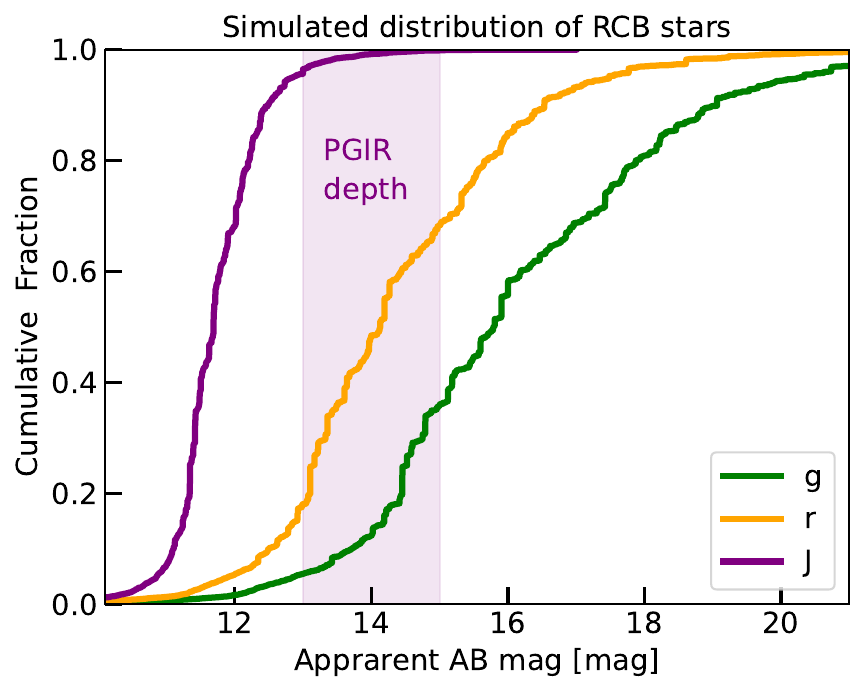}
    \caption{Distribution of the apparent magnitudes of simulated Galactic RCB stars at maximum light, that are visible from the northern hemisphere. The shaded purple region shows the typical range of PGIR limiting magnitudes (which varies depending on the extent of confusion). PGIR is sensitive enough to detect $>90\%$ of RCB stars at maximum light. PGIR can detect almost all RCB stars that brightened to maximum light in the last five years. PGIR has the same RCB-detection efficiency as an optical survey with a much deeper limiting magnitude ($\approx20$ mag) -- illustrating the advantages of a NIR search for dusty RCB stars.}
    
    \label{fig:rcb_simulated}
\end{figure}

Finally, we quantify the efficiency of PGIR in detecting RCB stars. We created a simulated distribution of 10000 RCB stars in the Milky Way assuming that the distribution traces the Galactic stellar mass. We use the SED of the prototype star R\,CrB together with interstellar extinction values from \citet{Green2019} to predict the expected distribution of the brightness of RCB stars visible from the northern hemisphere ($\delta>-28 ^{\rm{o}}$). Fig. \ref{fig:rcb_simulated} shows the  distribution of apparent magnitudes of the simulated RCB stars in the \emph{J} and optical \emph{g} and \emph{r} bands. We find that PGIR with a limiting magnitude m$_{\rm{lim}}\approx13-15$ mag is sensitive enough to detect $>90\%$ of RCB stars at maximum light. PGIR will thus detect almost all RCB stars in the northern hemisphere that brightened to maximum light over its five year baseline. Declines longer than 5 years are seen in known RCB stars, such as the historic 10-year dimming of R CrB itself in 2007. Examining declines in all known RCB stars using long-baseline AAVSO lightcurves, Crawford et al. (2024, in prep.) finds that 30 out of 1039 (~3\%) declines are longer than 5 years, and additionally that the coldest RCB stars are prone to both more frequent declines and spending more than 80\% of observed time in decline phase. These cool stars occasionally not show any brightness variations and would thus not be identified as large amplitude variables. However, they would still pass the lightcurve-independent color-selection criteria. Specifically, as mentioned in Sec. 5.1.1, col-pri 3-b accounts for highly dust-enshrouded cold RCB stars that spend most of their time in a dust-enshrouded phase and do not rise to maximum light
Interestingly, Fig. \ref{fig:rcb_simulated} shows that PGIR has the same RCB-detection efficiency as an optical survey with a much deeper limiting magnitude (m$_{\rm{lim}}\approx20$ mag, e.g. ZTF) -- illustrating the advantages of a NIR search at finding dusty RCB stars in dusty regions of the Milky Way. 

To summarize, we determine that there are a total of 350 RCB stars in the Milky Way (C.I. 250--500). Despite the biases associated with our search listed here, the total number of Galactic RCB stars is unlikely to be substantially larger than this estimate. 

\subsubsection{Comparisons with the rate of He-CO WD mergers}
Using our derived number of 350 (C.I. 250--500) Galactic RCB stars and assuming typical RCB lifetimes of $1-3 \times 10^{5}$\,yr \citep{Schwab2019, Crawford2020, Wong2024}, the formation rate of RCB stars,  in the Milky Way is between $0.8-5\times10^{-3}$\,yr$^{-1}$. This is consistent with observational and theoretical estimates of the He-CO WD merger rate. Using observations of low-mass WD binaries in the Milky Way disk from the ELM survey, \citet{Brown2020} estimate a lower limit of 2$\times10^{-3}$\,yr$^{-1}$ on the rate of He-CO WD mergers. Using binary population synthesis models, \citet{Karakas2015} estimate a He-CO WD merger rate of $\sim 1.8\times10^{-3}$ yr$^{-1}$. 

The population of RCB stars can provide useful information for future gravitational wave missions like LISA. Recently, \citet{Lamberts2019} found that close He-CO double white dwarfs (DWDs) will constitute the majority of sources detectable with LISA, and predicted over 5000 He-CO DWD resolvable over a 4 year baseline. As RCB stars are remnants of He-CO WD mergers, they serve as an independent probe of these predictions. We start with the simulated short-period (lower than few hours) He-CO WD binaries from \citet{Lamberts2019} and evolve them assuming gravitational wave radiation dominates the binary evolution, and find a He-CO WD merger rate of $\approx 1\times10^{-3}$\,yr$^{-1}$. Assuming RCB-lifetimes as above, this corresponds to an expected number of 100--300 RCB stars in the Milky Way. It is encouraging that this estimate is broadly consistent with the observed number of RCB stars despite simplified assumptions about WD binary evolution. More detailed simulations that account for different RCB-lifetimes, RCB progenitor-mass ranges and WD-merger physics such as mass transfer can provide observationally-grounded predictions for the dominant population of DWDs that should be detectable with LISA.

Missing from this picture are dLHdC stars. As noted in \citet{Tisserand2022}, there could be as many, if not more, dLHdC stars in the Milky Way as RCB stars. There are several indications that the population of WDs that merge to from RCB stars have distinct properties from those that form dLHdC stars \citep{Karambelkar2022, Tisserand2022, Crawford2022}. The lower luminosities and oxygen isotope ratios suggest that dLHdC stars could come from lower mass mergers than RCB stars. From BPS simulations, \citet{Tisserand2022} find that the distribution of total masses of WD merger remnants is bimodal, with the higher end ($\approx0.9$\,M$_{\odot}$) coming from hybrid-CO + CO WD mergers. If RCB stars form preferentially from these higher mass mergers, the RCB formation rate derived here corresponds to the rate of hybrid-CO + CO WD mergers rather than the full rate of He-CO WD mergers. Accurate mass measurements of RCB and dLHdC stars will help understand the implications of the RCB formation rate on the rate of WD mergers.

\subsection{Pulsation periods}
\label{sec:pulsation_periods}
At maximum light, some RCB stars are known to pulsate with periods between 40-100 days and amplitudes of a few tenths of a magnitude  \citep{Lawson1997, Alcock2001, Percy2023}. These pulsations can be fairly irregular $-$ the star can exhibit multiple pulsation modes or undergo changes in the dominant period (e.g. R CrB has shown pulsations with periods of 33, 44, 52 and 60 days \citep{Lawson1996}). Initially, these semi-regular or irregular pulsations were suggested to originate from the strange-mode instability in non-adiabatic and radiation-pressure dominated envelopes \citep{Saio2008, Gautschy2023}.  Recently, \citet{Wong2024} modeled the RCB-dLHdC pulsations as solar-like oscillations excited by convection in a helium-rich envelope. They find that the frequencies with maximum power ($\nu_{\rm{max}}$) for such oscillations matches the observed range of periods in RCBs and dLHdCs. These models show that the pulsation periods can be used as diagnostics of the mass of the stars, with lower periods generally indicative of lower masses. 
\begin{figure*}
    \includegraphics[width=\textwidth]{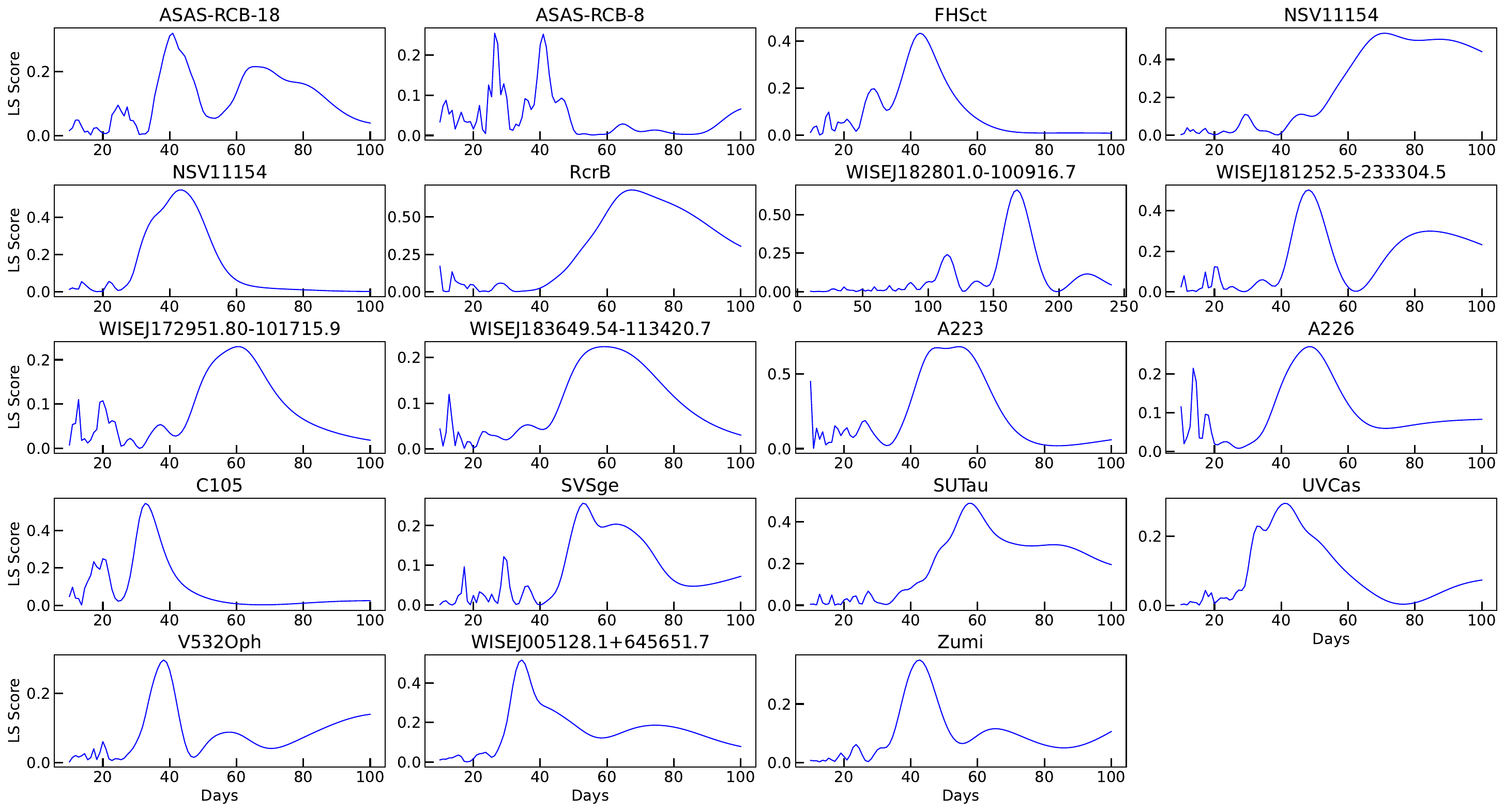}
    \caption{Periodograms of RCB and dLHdC stars using PGIR lightcurves reported in Table \ref{tab:rcb_pulsation_periods}}
\end{figure*}

We examined the \emph{J}-band PGIR light curves of the RCB stars presented in this paper, 70 previously known RCB stars to search for maximum-light periodic variations. We identified windows in the light curves that show periodic variations, and derived the periods using the Lomb-Scargle implementation in the \texttt{python} package \texttt{gatspy}. We derive pulsation periods for 16 RCB. To derive errorbars on the periods, for each star, we used the measured brightness and uncertainties to generate 100 simulations of the lightcurve assuming a normal distribution and measured the periods for each lightcurve. The median periods and standard deviations are listed in Table \ref{tab:rcb_pulsation_periods}. The periods typically lie between 30--100 days. Periods of some of these stars have been previously reported based on optical photometry -- SU\,Tau ($\sim 40$\,d, \citealp{Lawson1990}), FH\,Sct ($\sim$47\,d, \citealp{Percy2023}), R CrB (multiple periods ranging from $\sim35-60$\,d \citealp{Lawson1996}), C105 ($\sim 40$\,d, \citealp{Tisserand2022}). The PGIR periods of these stars are consistent with those reported previously.


\begingroup
\renewcommand{\tabcolsep}{1pt}
\begin{table}
\begin{center}
\begin{minipage}{8cm}
\caption{Pulsation periods using PGIR lightcurves}
\label{tab:rcb_pulsation_periods}
\begin{tabular}{lc}
\hline
\hline
{Name} & {PGIR Period} \\
{}     & {days} \\
\hline
ASAS-RCB-8       & 27 $\pm$ 6 \\
ASAS-RCB-18      & 41 $\pm$ 8 \\
FH Sct            &  44$\pm$ 11\\
NSV 11154         & 70 $\pm$ 20, 45 $\pm$ 5 \\
R CrB             & 67 $\pm$ 10 \\
SU Tau           & 58 $\pm$ 10 \\
SV Sge           & 53 $\pm$ 10 \\
UV Cas            & 41 $\pm$ 10  \\
V532 Oph          & 38 $\pm$ 10 \\

WISE J005128.09+645651.73           & 34 $\pm$ 6 \\
WISE J182801.05-100916.71           & 170 $\pm$ 10 \\
WISE J181252.50-233304.47           & 48 $\pm$ 15 \\
WISE J172951.80-101715.9  & 60 $\pm$ 19 \\
WISE J183649.54-113420.7  & 59 $\pm$ 9 \\
Z Umi             & 42 $\pm$ 10 \\
C105  & 33 $\pm$ 10\\

\hline
\hline
\end{tabular}
\end{minipage}
\end{center}
\end{table}
\endgroup



In addition to the PGIR lightcurves, we also examined lightcurves of RCB and dLHdC stars from the Transiting Exoplanet Satellite Survey (\emph{TESS}, \citealp{Ricker2015}) to search for short-timescale variability from them. We downloaded calibrated, short-cadence \emph{TESS} lightcurves from the Quicklook Pipeline (QLP) for known RCB and dLHdC stars from the Mikulski Archive for Space Telescopes \footnote{\href{https://mast.stsci.edu}{https://mast/stsci.edu}} using astroquery. We examined simple aperture photometry (SAP) lightcurves for each star and removed lightcurves which have bad photometric flags, large photometric scatter and rapid increases or decreases of flux that are likely not astrophysical. Additionally, we also use the python package \texttt{lightkurve} to download the \emph{TESS} target pixel files and extract aperture photometry at the locations of known RCB and dLHdC stars. We find that the QLP lightcurves agree well with those extracted using \texttt{lightkurve}, where available. We use the DREAMS-RCB monitoring website to determine the photometric phase of the RCB stars at the time of \emph{TESS} observations and use only those lightcurves that were observed at maximum light. We are then left with 6 RCB stars and 6 dLHdC stars. The \emph{TESS} lightcurves have cadences of 10 and 30 minutes, and a baseline of $\approx 22-27$ days per sector. Some stars were observed in multiple sectors. As it is challenging to stitch data from different sectors together if they are not observed continuously, we analyze the different sectors individually. 

\begin{figure*}[hbt]
    \includegraphics[width=0.5\textwidth]{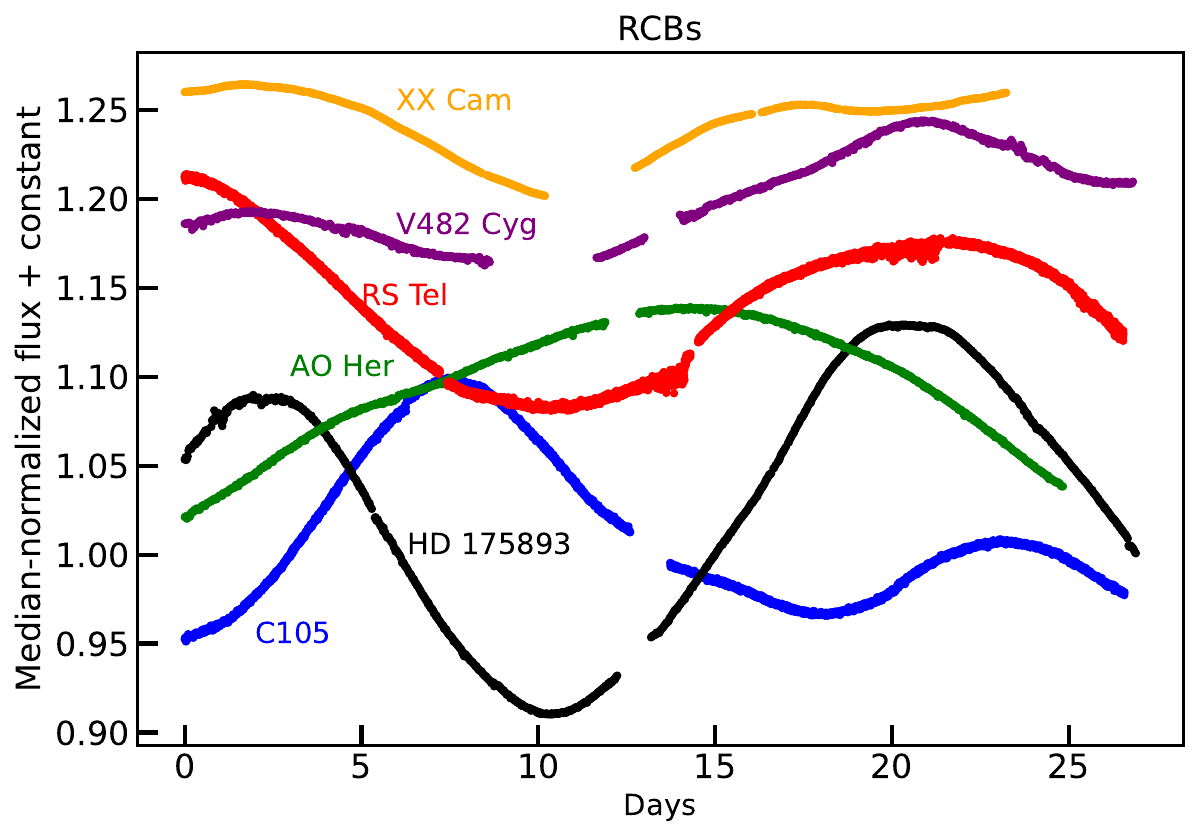}
    \includegraphics[width=0.5\textwidth]{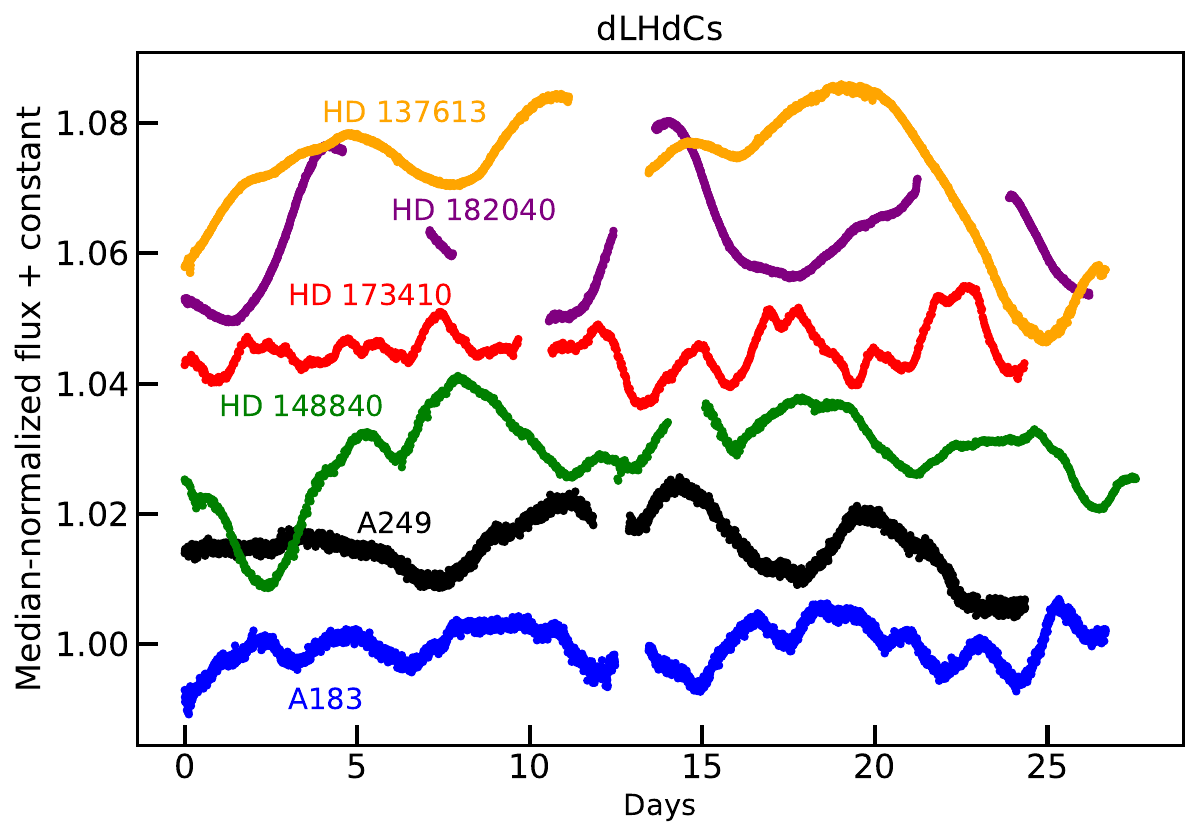}\\
    \caption{\emph{Top : }Examples of \emph{TESS} lightcurves of RCB stars (\emph{left}) and dLHdC stars (\emph{right}). \emph{Bottom :} Lomb-Scargle periodograms of RCB stars (\emph{left}) and dLHdC stars (\emph{right}). dLHdC stars in general show variability on timescales shorter than RCB stars. A possible explanation is that dLHdC stars have lower mass than RCB stars.}
    \label{fig:tess_lcs}
\end{figure*}

Figure \ref{fig:tess_lcs} shows the median-normalized \emph{TESS} lightcurves of six RCB and six dLHdC stars. The \emph{TESS} lightcurves show that both RCB and dLHdC stars do not show variability at very short ($< 1$ day) timescales. Interestingly, these high-cadence lightcurves show that dLHdC stars generally exhibit variability on timescales shorter than RCB stars.  In the picture that dLHdC stars have lower masses than RCB stars, they are expected to have smaller pulsation periods than RCB stars \citep{Wong2024}. The differences seen in the \emph{TESS} lightcurves of RCBs and dLHdCs may therefore point towards the different masses of this class of objects, as suggested in \citet{Tisserand2022, Karambelkar2022}. Studying the variations of a larger sample of dLHdC and RCB stars will provide an important clue towards identifying their masses and possible differences. 

\subsection{Radial velocities}
We use our NIR spectra to derive the radial velocities (RVs) of the newly identified RCB stars. We correct the spectra for barycentric motion, examine them to identify strong carbon absorption lines, and fit a gaussian profile to derive the line centers. We then compare the line-centers to rest-wavelengths (taken from NIST) to measure the RVs. For each spectrum, we determined the statistical uncertainty on the RV using by calculating the standard deviation of the velocities measured from each carbon line examined. These uncertainties are small (typically $< 5$\,km\,s$^{-1}$). To determine the systematic uncertainty on how precisely line centroids can be measured due to instrumental resolution, we use the sky-emission lines in our spectra. The dispersion on the sky line centroids is ~20 km\,s$^{-1}$ (approx. a third of a pixel). We add the statistical and systematic uncertainties in quadrature and report those as uncertainties on the RVs in Table \ref{tab:rcb_radial_vels}.

\begingroup
\renewcommand{\tabcolsep}{3pt}
\begin{table}
\begin{minipage}{8cm}
\caption{Radial velocities derived from C absorption lines in the NIR spectra. Sources marked with * denote revised values from the ones reported previously in \citet{Karambelkar2021}}
\label{tab:rcb_radial_vels}
\begin{tabular}{lccc}
\hline
\hline
{Name} & {Radial velocity} \\
{}     & {km/s} \\
\hline
WISE-ToI-164 & $-27 \pm  20$ \\
WISE-ToI-174 & $-96 \pm  25$ \\
WISE-ToI-185 & $-68 \pm  20$ \\
WISE-ToI-188 & $-9 \pm  20$ \\
WISE-ToI-193 & $-122 \pm  20$ \\
WISE-ToI-194 & $120 \pm  20$ \\
WISE-ToI-207 & $-40 \pm  20$ \\
WISE-ToI-220 & $-101 \pm  20$ \\
WISE-ToI-226 & $-146 \pm  20$ \\
WISE-ToI-231 & $7 \pm  25$ \\
WISE-ToI-248 & $-44 \pm  25$ \\
WISE-ToI-257 & $27 \pm  20$ \\
WISE-ToI-264 & $120 \pm  20$ \\
WISE-ToI-270 & $69 \pm  22$ \\
WISE-ToI-323 & $-119 \pm  20$ \\
WISE-ToI-1227 & $-43 \pm  20$ \\
WISE-ToI-1241 & $-74 \pm  20$ \\
WISE-ToI-1245 & $3 \pm  20$ \\
WISE-ToI-2645 & $40 \pm  20$ \\
WISE-ToI-2938 & $-39 \pm  20$ \\
WISE-ToI-4108 & $173 \pm  20$ \\
WISE-ToI-4117 & $18 \pm  20$ \\
*AO Her & $-524 \pm  20$ \\ 
*ASAS-RCB-21 & $20 \pm  20$ \\ 
*NSV 11154 & $-326 \pm  22$ \\ 
*V391 Sct & $-36 \pm  25$ \\ 
*WISE-ToI-249$^{a}$ & $81 \pm  20$ \\ 
*WISE-ToI-203$^{a}$ & $-18 \pm  20$ \\ 
*WISE-ToI-290$^{a}$  & no C lines\\ 
*WISE-ToI-6 & $-101 \pm  20$ \\ 
*WISE-ToI-223 & $37 \pm  20$ \\
*WISE-ToI-268 & $95 \pm  20$ \\
*WISE-ToI-274 & $-36 \pm  20$ \\
*WISE-ToI-1309 & $-13 \pm  20$ \\
*WISE-ToI-281  & no C lines\\
*WISE-ToI-1213 & no C lines\\

\hline
\hline
\end{tabular}
\begin{tablenotes} 
\item $a$ : The stars WISE-ToI-203, 249 and 290 are listed as WISE-J17+, WISE-J18+ and WISE-J19+ in \citet{Karambelkar2021}.
\end{tablenotes}

\end{minipage}
\end{table}
\endgroup

We also note an error in the RVs reported in our previous paper  \citep{Karambelkar2021} -- the signs of the values in Table 3 there should be flipped. Additionally, we note that the previous values were measured by cross-correlating the entire spectrum with synthetic RCB spectra, while here, we measure the values using only strong carbon absorption lines. We find that the RVs measured using carbon lines are more reliable, as the spectral features in the synthetic models are highly dependent on the assumed elemental abundances of the RCB star. Indeed, we find that the RVs reported here agree better with those reported by other sources wherever available (e.g. \emph{Gaia}, see \citealt{Tisserand2023b}). Table \ref{tab:rcb_radial_vels} also lists the revised RVs for these stars. Most RCB stars in Table \ref{tab:rcb_radial_vels} have low RVs ($\leq 50$\,km\,s$^{-1}$), consistent with other RCB stars with RV measurements \citep{Tisserand2023b}. AO\,Her and NSV\,11154 have very high RVs, and are located towards the Galactic halo. The RVs reported here will be useful in constructing the 3D-distribution of Galactic RCB stars (similar to \citealt{Tisserand2023b}).

\subsection{Are RCB and DY\,Per type stars related?}
First, we note that neither DY\,Per, nor the three spectroscopically confirmed DY\,Per type stars show the He\,I ($\lambda 10830$) line that is ubiquitous in RCB stars. In RCB stars, this line is collisionally excited in high velocity ($\approx 400$\,km\,s$^{-1}$) He-rich dust-driven winds \citep{Clayton2013, Karambelkar2021}. DY\,Per type stars resemble classical carbon-stars in this aspect. However, even in the cold-RCB picture, the absence of this line can possibly be explained by the low-luminosity of DY\,Per type stars compared to RCB stars. As DY\,Per type stars are $\approx 10$ times dimmer than RCBs, the radiation pressure can accelerate only low velocity winds ($\approx40$\,km\,s$^{-1}$, assuming ), which is not sufficient to excite the helium atoms to the lower energy level of the He\,I transition. 

Second, Fig. \ref{fig:rcb_dyper_colcol_dist} (bottom panel) shows the Galactic distribution of the RCB and DY\,Per type stars identified in this paper together with known RCB and DY\,Per type stars. Most RCB stars lie towards the Galactic center, with a small number at higher Galactic latitudes suggestive of a small halo population. In contrast, the DY\,Per type stars and candidates lie at high Galactic longitudes, suggesting that they are part of a disk population. This is consistent with the findings of \citet{Tisserand2023b} who studied the distribution of RCB and DY\,Per type stars using \emph{Gaia} DR3. The top panel of Fig. \ref{fig:dyper_cands_lcs} shows the NIR color-color diagram for RCB and DY\,Per type stars. We find that the new DY\,Per type stars and DY\,Per candidates occupy a distinct region in this diagram from RCB stars and have colors similar to classical carbon stars, consistent with \citet{Tisserand2013}. 

We also note that the new DY\,Per type stars and candidates show some diversity in their lightcurves. We plot longer baseline ATLAS-o band lightcurves of the DY\,Per type stars and candidates in Fig. \ref{fig:dyper_lcs} and Fig. \ref{fig:dyper_cands_lcs}. Some stars (e.g. PGIRV\_396, PGIRV\_230\_1\_0\_3575, NIKC\,2-77, V*FL\,Per, Fuen\,C\,157) show well-defined brightness-declines that are clearly distinguishable from other small-amplitude variations, while some stars (e.g. IRAS\,04193+, IRAS\,07113+, C*2905, Fig. \ref{fig:dyper_cands_lcs}) generally show large-amplitude erratic variations without well-defined declines. The lightcurves of some stars such as PGIRV\_230\_2\_0\_290 (Fig. \ref{fig:dyper_lcs}), C\*2905 and ATO\,J308.8118 (Fig. \ref{fig:dyper_cands_lcs}) show large-amplitude pulsations that are seen in carbon stars. The stars V2060\,Cyg, KISO\,C1-139, IRAS22137+ (Fig. \ref{fig:dyper_cands_lcs}) show declines at periodic intervals. Based on the lightcurves of classical carbon stars in the LMC, \citep{Soszynski2009} suggest that DY\,Per variability is part of the continuum of carbon-star variability. Spectroscopic observations of the different classes of DY\,Per candidates will help understand which, if any, of these stars show RCB-like elemental abundances (esp. $^{18}$O).

\section{Summary and way forward}
\label{sec:summary}
In this paper, we presented results from a systematic infrared census for RCB stars in the Milky Way. We selected RCB candidates using NIR \emph{J-}band lightcurves from PGIR, mid-IR colors from WISE and obtained medium resolution NIR spectra for them. We identified 53 RCB stars from our candidates. We use this number to estimate the total number of RCB stars in the Milky Way. This has been a longstanding open question - with estimates ranging from a few thousand \citep{Clayton12, Han98, Alcock2001} to a few hundred \citep{Tisserand2020}. Our systematic infrared census provides an excellent way to address this question. Using our selection criteria, we estimate that there are a total of $\approx$\,350 RCB stars with a 95\% confidence interval of 250 -- 500 in the Milky Way. This corresponds to a formation rate of $0.8-5\times10^{-3}$\,yr$^{-1}$. This is consistent with observational and theoretical estimates of the rate of He-CO WD mergers in the Milky Way. Using binary population synthesis models, the measured RCB-formation rate can be used to draw insights about the population of He-CO WD binaries detectable with future gravitational-wave experiments such as \emph{LISA}. However, in addition to RCB stars, it is important to understand the contribution of the dustless dLHdC stars and colder DY\,Per type stars to the population of He-CO WD merger remnants. 

It is still not clear whether DY\,Per type stars are colder RCB stars or classical carbon stars. Only 3 Galactic DY\,Per type stars were known in the Milky Way. In this paper, we identified 3 spectroscopically confirmed and 15 candidate DY\,Per type stars. The new DY\,Per type stars and candidates have distinct NIR colors and appear to have a different Galactic distribution than RCB stars. Future analysis of the spectra and long-term photometric variations of these stars will be useful to understand their relation to RCB stars. dLHdC stars have been conclusively associated with He-CO WD merger remnants, but their number is uncertain. As noted in \citep{Tisserand2022}, there could potentially be as many dLHdC stars as RCB stars in the Milky Way. A systematic search for dLHdC stars is required to interpret the number of RCB-dLHdC stars in the context of the He-CO WD merger rate.  

The differences between dLHdC and RCB stars have only recently started to be explored. The differences in their luminosities and chemical compositions suggest that dLHdC stars could be less massive than RCB stars \citep{Tisserand2022, Karambelkar2022, Crawford2022}. In this picture, we would expect the maximum-light pulsation periods of these stars to differ. The \emph{TESS} lightcurves for six dLHdC and six RCB stars show that dLHdC stars show variations on timescales shorter than RCB stars, consistent with the picture that they have lower masses. Comparing these pulsation data to theoretical models (e.g. \citealp{Wong2024, Saio2008}) to estimate their masses will providing useful information towards understand why dLHdCs form dust and RCBs do not.

Finally, we have presented NIR spectra for 44 RCB stars, which can be useful for measuring their elemental abundances. In addition to oxygen isotope ratios (e.g. \citealp{Karambelkar2021, Karambelkar2022}), it would be interesting to see if the NIR spectra can be used to solve the long-standing carbon problem in RCBs \citep{Asplund2000}. NIR spectra also probe the helium line, which can be used to study mass-loss in RCB stars \citep{Clayton2011}. 

NIR observations are an efficient way to identify and characterize RCB stars. Ongoing and upcoming NIR surveys such as the Wide-field Infrared Transient Explorer (WINTER, \citealt{Lourie2020}), the Dynamic Red All-sky Monitoring Survey (DREAMS), Cryoscope in Antarctica will help uncover the population of RCB stars in the Milky Way. Future missions such as the \emph{Nancy Grace Roman Space Telescope} will help study the RCB stars in the most crowded central region of the Milky Way. In the optical, the Vera Rubin Observatory has the sensitivity to discover all Galactic RCB stars in the southern hemisphere, as well as RCB stars in other galaxies out to $\approx$5\,Mpc, shedding light on the DWD populations of these galaxies.



\section*{Acknowledgements}
We thank the anonymous referee for useful suggestions that improved this paper. VK thanks Sunny Wong and Yashvi Sharma for useful discussions. Palomar Gattini-IR (PGIR) is generously funded by Caltech, Australian National University, the Mt Cuba Foundation, the Heising Simons Foundation, the Bi- national Science Foundation. PGIR is a collaborative project among Caltech, Australian National University, University of New South Wales, Columbia University and the Weizmann Institute of Science. MMK acknowledges generous support from the David and Lucille Packard Foundation. J. Soon acknowledges the support of an Australian Government Research Training Program (RTP) scholarship. Some of the data presented here were obtained with Visiting Astronomer facility at the Infrared Telescope Facility, which is operated by the University of Hawaii under contract 80HQTR19D0030 with the National Aeronautics and Space Administration. This work was supported, in part, by the National Science Foundation through grant PHY-2309135 to the Kavli Institute for Theoretical Physics, and by the Gordon and Betty Moore Foundation through grant GBMF5076. 

\section*{Data Availability}
The PGIR \emph{J-}band lightcurves of all 1215 candidates, lightcurves of new RCB stars, NIR spectra of all 453 sources, spectroscopic classifications of all 453 sources (Table \ref{tab:classification_catalog}) and updated priorities of the T20 candidates (Table \ref{tab:updated_priorities}) are publicly available at Zenodo at \href{10.5281/zenodo.12683154}{10.5281/zenodo.12683154}.

\appendix
\section{Candidate RCB and DY\,Per type stars}
\begingroup
\renewcommand{\tabcolsep}{8pt}
\begin{table*}
\begin{center}
\begin{minipage}{18cm}
\caption{Strong RCB and DY\,Per candidates identified from our NIR census.}
\label{tab:rcbs_cands_table}
\begin{tabular}{lccccc}
\hline
\hline
{Name} & {ToI--ID/} & {RA} & {Dec} & {NIR spec.} & {Comments} \\
{}     &  {PGIR Name}      & {deg} & {deg} & {class} & {}  \\
\hline
\hline
WISE J060405.01+233304.7 &   28 &  91.02088 & 23.55132 & RCB-cand & \\ 
WISE J072356.66-124014.0 &   41 & 110.98611 & -12.67058 & RCB-cand & \\ 
WISE J182649.64-244532.8 &  228 & 276.70684 & -24.75912 & RCB-cand & \\ 
WISE J184016.09-035608.9 &  245 & 280.06707 & -3.93583 & 
RCB-cand & \\
WISE J193929.35+244504.0 &  288 & 294.87231 & 24.75113 & RCB-cand & \\ 
WISE J194739.93+232638.7 &  293 & 296.91639 & 23.44410 & RCB-cand & \\ 
WISE J222704.54-165948.4 &  323 & 336.76892 & -16.99678 & RCB-cand & \\ 
WISE J181400.05-134254.4 & 1257 & 273.50025 & -13.71511 & RCB-cand & \\ 
\hline
Kiso C1-139 & & 18.13093 & 62.18622 & DY Per-cand & \textbf{\citet{Maehara1987}}\\
NIKC 2-77 & & 55.08234 & 59.09781 & DY Per-cand & \textbf{\citet{Soyano1991}}\\
Fuen C 157 & & 90.42725 & 24.56634 & DY Per-cand & \textbf{\citet{Fuenmayor1981}}\\
V* FL Per & & 59.90524 & 46.46239 & DY Per-cand & \textbf{\citet{Lee1947}}\\
Case 492 & & 330.83864 & 62.30802 & DY Per-cand & \textbf{\citet{Nassau1957}} \\
IRAS 22137+6311 & & 333.83008 & 63.44256 & DY Per-cand & \textbf{\citet{Alksnis2001}}\\
IRAS 07113-0025 & & 108.47103 & -0.51654 & DY Per-cand & \textbf{\citet{Alksnis2001}}\\
V* AR Vul & & 293.93229 & 26.55986 & DY Per-cand & \textbf{\citet{Nassau1957}}\\
IRAS 04193+4959 & & 65.78151 & 50.10781 & DY Per-cand & \textbf{\citet{Putney1997}}\\
V* V2060 Cyg & & 317.43137 & 54.17533 & DY Per-cand & \textbf{\citet{Alksnis2001}}\\
Kiso C1-24 & & 357.80148 & 62.38434 & DY Per-cand & \textbf{\citet{Maehara1987}}\\
ABC90 cep 7 & & 335.68037 & 54.59463 & DY Per-cand & \textbf{\citet{Alksnis2001}}\\
Case 749 & & 334.65993 & 43.77903 & DY Per-cand & \textbf{\citet{Blanco1958}}\\
C* 2905 & & 307.02683 & 42.91874 & DY Per-cand & \textbf{\citet{Stephenson1973}}\\
ATO J308.8118+45.2629 & & 308.81182 & 45.26292 & DY Per-cand & \textbf{\citet{Alksnis2001}}\\
\hline
\hline
\end{tabular}
\begin{tablenotes}
    \item \textbf{All strong RCB candidates are sources from the WISE color-selected catalog of \citet{Tisserand2020}. For the DY Per-type candidates, we list the papers that first classified them as carbon stars.}
\end{tablenotes}
\end{minipage}
\end{center}
\end{table*}
\endgroup

\begin{figure*}
    \centering
    \includegraphics[width=\textwidth]{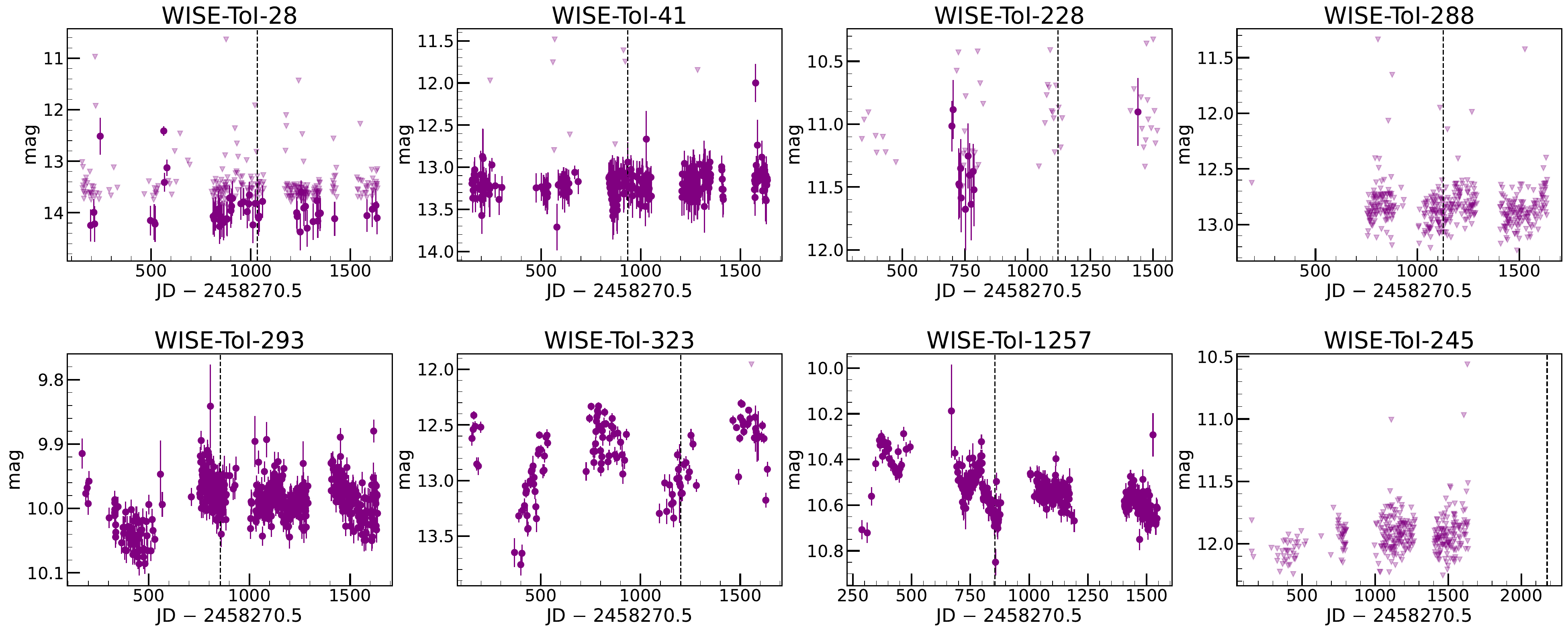}\\    \includegraphics[width=0.95\textwidth]{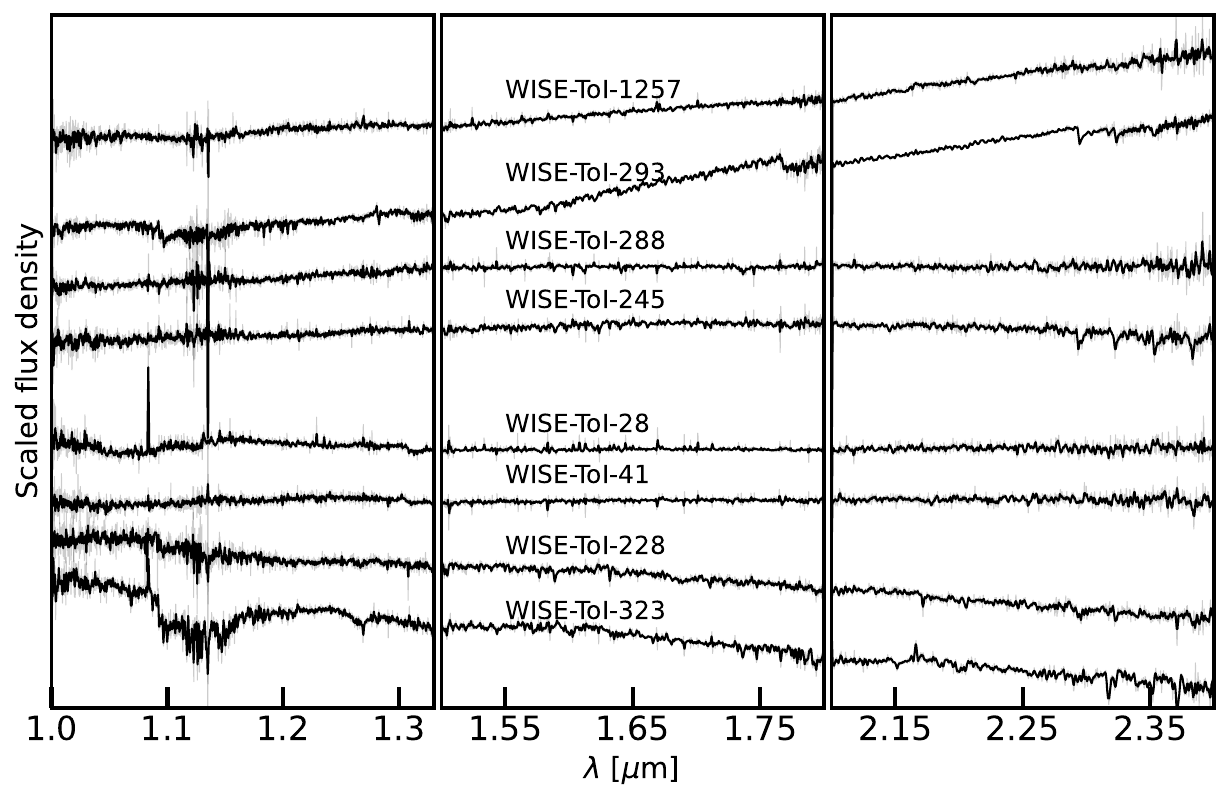}
    \caption{NIR spectra of the 8 candidate RCB stars listed in Table \ref{tab:rcbs_table}. The broad absorption feature seen in the spectrum of WISE-ToI-323 from 1.1 -- 1.2$\mu$m is likely due to imperfect telluric correction.}
    \label{fig:new_cand_rcbs_2_spectra}
\end{figure*}

\begin{figure*}
    \includegraphics[width=\textwidth]{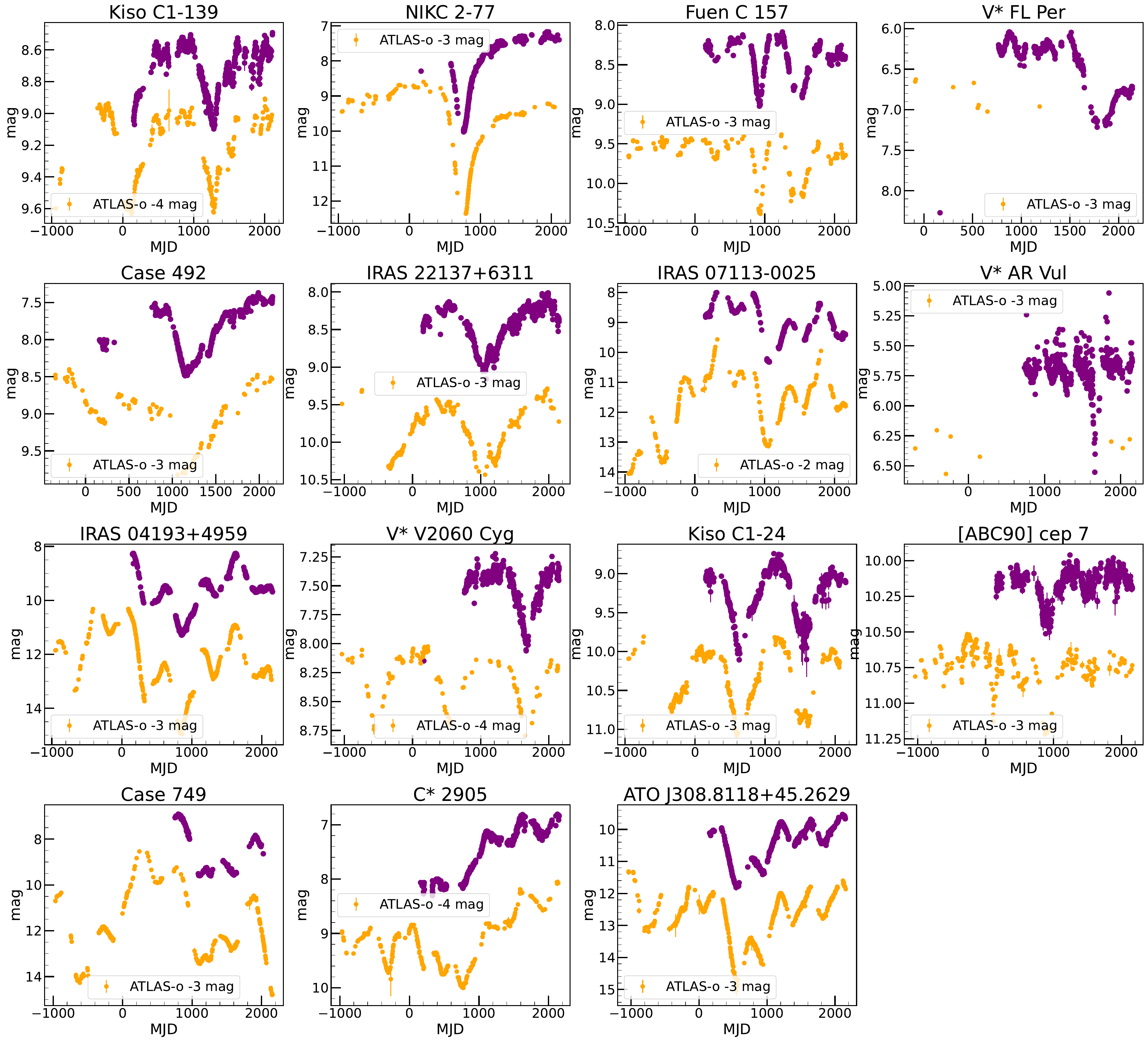}
    \caption{PGIR J-band lightcurves of candidate DY\,Per type stars}
    \label{fig:dyper_cands_lcs}
\end{figure*}

Table \ref{tab:rcbs_cands_table} lists the candidate RCB and DY\,Per type stars identified in our paper. Fig. \ref{fig:new_cand_rcbs_2_spectra} shows the lightcurves and spectra of the stars listed as strong RCB candidates. Fig. \ref{fig:dyper_cands_lcs} shows the lightcurves of stars listed as candidate DY\,Per type stars.

\section{Source classifications}
\label{appendix:classifications}
We now discuss classifications of the 453 spectra presented in this paper. 

\emph{M stars} -- A total of 154 stars have M-type spectra (with broad TiO, VO features) and are thus O-rich AGB stars. We compare the spectra to the IRTF spectral library \citep{Rayner2009}, and determine the best-fit match by performing a least-squares fit. The spectral types of these stars range between M5--M9. Several of these stars show H emission lines -- commonly seen in Miras. The strength of these emission lines are known to vary with pulsation phase, and likely originate in pulsation-driven shock heating of the atmospheres \citep{Gray2009}. 

\emph{Possible symbiotic binaries} -- 15 stars show AGB-star-like spectra together with strong emission lines, particularly the \ion{He}{1} $\lambda$10830 emission line. These stars are possibly symbiotic stars, where the helium line collisionally excited in high velocity winds around the star. 8 stars have spectra similar to M-type stars, while 7 stars have spectra resembling C stars. 

\emph{Emission stars} -- 32 stars show spectra dominated by emission lines, we classify them as emission stars.

\emph{Dust forming carbon stars} -- 42 stars show spectra resembling carbon-stars. 39 of these show a broad absorption feature at 1.5\um. This feature is likely due to HCN + C$_{2}$H$_{2}$ and has been previously noted in several carbon stars \citet{Gonneau2016, GautschyLoidl2004}. 

\emph{CO emitters : Possible RV-Tauri or young-stellar objects (YSOs)} -- 69 stars show CO emission bands in their spectra.  Some of these stars show short-period variations on top of a long-period, large amplitude variation, characteristic of RV-Tauri stars.  The remaining stars are likely a combination of RV-Tauri stars and young stellar objects (YSOs) that are known to exhibit CO emission in their spectra. 

\emph{H-rich} -- 71 stars have spectra with strong absorption lines of hydrogen. These stars are most likely RV-Tauri stars. 

\emph{Other} -- We identify 2 Wolf-Rayet and 1 post-AGB star. 9 additional stars do not show any obvious strong features in their spectra except some hydrogen lines and no brightness variations, suggesting that they are not RCB stars.



\bibliography{myreferences}

\begin{thebibliography}{}
\expandafter\ifx\csname natexlab\endcsname\relax\def\natexlab#1{#1}\fi

\bibitem[{{Alcock} {et~al.}(2001){Alcock}, {Allsman}, {Alves}, {Axelrod}, {Becker}, {Bennett}, {Clayton}, {Cook}, {Dalal}, {Drake}, {Freeman}, {Geha}, {Gordon}, {Griest}, {Kilkenny}, {Lehner}, {Marshall}, {Minniti}, {Misselt}, {Nelson}, {Peterson}, {Popowski}, {Pratt}, {Quinn}, {Stubbs}, {Sutherland }, {Tomaney}, {Vandehei}, \& {Welch}}]{Alcock2001}
{Alcock}, C., {Allsman}, R.~A., {Alves}, D.~R., {et~al.} 2001, \apj, 554, 298

\bibitem[{{Alksnis} {et~al.}(2001){Alksnis}, {Balklavs}, {Dzervitis}, {Eglitis}, {Paupers}, \& {Pundure}}]{Alksnis2001}
{Alksnis}, A., {Balklavs}, A., {Dzervitis}, U., {et~al.} 2001, Baltic Astronomy, 10, 1

\bibitem[{{Asplund} {et~al.}(2000){Asplund}, {Gustafsson}, {Lambert}, \& {Rao}}]{Asplund2000}
{Asplund}, M., {Gustafsson}, B., {Lambert}, D.~L., \& {Rao}, N.~K. 2000, \aap, 353, 287

\bibitem[{{Bailer-Jones} {et~al.}(2018){Bailer-Jones}, {Rybizki}, {Fouesneau}, {Mantelet}, \& {Andrae}}]{Bailer-Jones2018}
{Bailer-Jones}, C.~A.~L., {Rybizki}, J., {Fouesneau}, M., {Mantelet}, G., \& {Andrae}, R. 2018, \aj, 156, 58

\bibitem[{{Bellm} {et~al.}(2019){Bellm}, {Kulkarni}, {Graham}, {Dekany}, {Smith}, {Riddle}, {Masci}, {Helou}, {Prince}, {Adams}, {Barbarino}, {Barlow}, {Bauer}, {Beck}, {Belicki}, {Biswas}, {Blagorodnova}, {Bodewits}, {Bolin}, {Brinnel}, {Brooke}, {Bue}, {Bulla}, {Burruss}, {Cenko}, {Chang}, {Connolly}, {Coughlin}, {Cromer}, {Cunningham}, {De}, {Delacroix}, {Desai}, {Duev}, {Eadie}, {Farnham}, {Feeney}, {Feindt}, {Flynn}, {Franckowiak}, {Frederick}, {Fremling}, {Gal-Yam}, {Gezari}, {Giomi}, {Goldstein}, {Golkhou}, {Goobar}, {Groom}, {Hacopians}, {Hale}, {Henning}, {Ho}, {Hover}, {Howell}, {Hung}, {Huppenkothen}, {Imel}, {Ip}, {Ivezi{\'c}}, {Jackson}, {Jones}, {Juric}, {Kasliwal}, {Kaspi}, {Kaye}, {Kelley}, {Kowalski}, {Kramer}, {Kupfer}, {Landry}, {Laher}, {Lee}, {Lin}, {Lin}, {Lunnan}, {Giomi}, {Mahabal}, {Mao}, {Miller}, {Monkewitz}, {Murphy}, {Ngeow}, {Nordin}, {Nugent}, {Ofek}, {Patterson}, {Penprase}, {Porter}, {Rauch}, {Rebbapragada}, {Reiley}, {Rigault}, {Rodriguez}, {van Roestel}, {Rusholme}, {van
  Santen}, {Schulze}, {Shupe}, {Singer}, {Soumagnac}, {Stein}, {Surace}, {Sollerman}, {Szkody}, {Taddia}, {Terek}, {Van Sistine}, {van Velzen}, {Vestrand}, {Walters}, {Ward}, {Ye}, {Yu}, {Yan}, \& {Zolkower}}]{Bellm2019}
{Bellm}, E.~C., {Kulkarni}, S.~R., {Graham}, M.~J., {et~al.} 2019, \pasp, 131, 018002

\bibitem[{{Bhowmick} {et~al.}(2018){Bhowmick}, {Pandey}, {Joshi}, \& {Ashok}}]{Bhowmick2018}
{Bhowmick}, A., {Pandey}, G., {Joshi}, V., \& {Ashok}, N.~M. 2018, \apj, 854, 140

\bibitem[{{Blanco}(1958)}]{Blanco1958}
{Blanco}, V.~M. 1958, \apj, 127, 191

\bibitem[{Brown {et~al.}(2020)Brown, Kilic, Kosakowski, Andrews, Heinke, Agüeros, Camilo, Gianninas, Hermes, \& Kenyon}]{Brown2020}
Brown, W.~R., Kilic, M., Kosakowski, A., {et~al.} 2020, The Astrophysical Journal, 889, 49

\bibitem[{{Clayton}(1996)}]{Clayton1996}
{Clayton}, G.~C. 1996, \pasp, 108, 225

\bibitem[{{Clayton}(2012)}]{Clayton12}
---. 2012, Journal of the American Association of Variable Star Observers (JAAVSO), 40, 539

\bibitem[{{Clayton} {et~al.}(2007){Clayton}, {Geballe}, {Herwig}, {Fryer}, \& {Asplund}}]{Clayton2007}
{Clayton}, G.~C., {Geballe}, T.~R., {Herwig}, F., {Fryer}, C., \& {Asplund}, M. 2007, \apj, 662, 1220

\bibitem[{{Clayton} {et~al.}(2013){Clayton}, {Geballe}, \& {Zhang}}]{Clayton2013}
{Clayton}, G.~C., {Geballe}, T.~R., \& {Zhang}, W. 2013, \aj, 146, 23

\bibitem[{{Clayton} {et~al.}(2005){Clayton}, {Herwig}, {Geballe}, {Asplund}, {Tenenbaum}, {Engelbracht}, \& {Gordon}}]{Clayton2005}
{Clayton}, G.~C., {Herwig}, F., {Geballe}, T.~R., {et~al.} 2005, \apjl, 623, L141

\bibitem[{{Clayton} {et~al.}(1992){Clayton}, {Whitney}, {Stanford}, \& {Drilling}}]{Clayton1992}
{Clayton}, G.~C., {Whitney}, B.~A., {Stanford}, S.~A., \& {Drilling}, J.~S. 1992, \apj, 397, 652

\bibitem[{{Clayton} {et~al.}(2011){Clayton}, {Sugerman}, {Stanford}, {Whitney}, {Honor}, {Babler}, {Barlow}, {Gordon}, {Andrews}, {Geballe}, {Bond}, {De Marco}, {Lawson}, {Sibthorpe}, {Olofsson}, {Polehampton}, {Gomez}, {Matsuura}, {Hargrave}, {Ivison}, {Wesson}, {Leeks}, {Swinyard}, \& {Lim}}]{Clayton2011}
{Clayton}, G.~C., {Sugerman}, B. E.~K., {Stanford}, S.~A., {et~al.} 2011, \apj, 743, 44

\bibitem[{{Crawford} {et~al.}(2020){Crawford}, {Clayton}, {Munson}, {Chatzopoulos}, \& {Frank}}]{Crawford2020}
{Crawford}, C.~L., {Clayton}, G.~C., {Munson}, B., {Chatzopoulos}, E., \& {Frank}, J. 2020, \mnras, 498, 2912

\bibitem[{{Crawford} {et~al.}(2022){Crawford}, {Tisserand}, {Clayton}, \& {Munson}}]{Crawford2022}
{Crawford}, C.~L., {Tisserand}, P., {Clayton}, G.~C., \& {Munson}, B. 2022, \aap, 667, A85

\bibitem[{{Crawford} {et~al.}(2023){Crawford}, {Tisserand}, {Clayton}, {Soon}, {Bessell}, {Wood}, {Garc{\'\i}a-Hern{\'a}ndez}, {Ruiter}, \& {Seitenzahl}}]{Crawford2023}
{Crawford}, C.~L., {Tisserand}, P., {Clayton}, G.~C., {et~al.} 2023, \mnras, 521, 1674

\bibitem[{{Cushing} {et~al.}(2004){Cushing}, {Vacca}, \& {Rayner}}]{Cushing2004}
{Cushing}, M.~C., {Vacca}, W.~D., \& {Rayner}, J.~T. 2004, \pasp, 116, 362

\bibitem[{{Czekaj} {et~al.}(2014){Czekaj}, {Robin}, {Figueras}, {Luri}, \& {Haywood}}]{Czejak2014}
{Czekaj}, M.~A., {Robin}, A.~C., {Figueras}, F., {Luri}, X., \& {Haywood}, M. 2014, \aap, 564, A102

\bibitem[{{De} {et~al.}(2020){De}, {Hankins}, {Kasliwal}, {Moore}, {Ofek}, {Adams}, {Ashley}, {Babul}, {Bagdasaryan}, {Burdge}, {Burnham}, {Dekany}, {Declacroix}, {Galla}, {Greffe}, {Hale}, {Jencson}, {Lau}, {Mahabal}, {McKenna}, {Sharma}, {Shopbell}, {Smith}, {Soon}, {Sokoloski}, {Soria}, \& {Travouillon}}]{De2020}
{De}, K., {Hankins}, M.~J., {Kasliwal}, M.~M., {et~al.} 2020, \pasp, 132, 025001

\bibitem[{{Eyer} {et~al.}(2023){Eyer}, {Audard}, {Holl}, {Rimoldini}, {Carnerero}, {Clementini}, {De Ridder}, {Distefano}, {Evans}, {Gavras}, {Gomel}, {Lebzelter}, {Marton}, {Mowlavi}, {Panahi}, {Ripepi}, {Wyrzykowski}, {Nienartowicz}, {Jevardat de Fombelle}, {Lecoeur-Taibi}, {Rohrbasser}, {Riello}, {Garc{\'\i}a-Lario}, {Lanzafame}, {Mazeh}, {Raiteri}, {Zucker}, {{\'A}brah{\'a}m}, {Aerts}, {Aguado}, {Anderson}, {Bashi}, {Binnenfeld}, {Faigler}, {Garofalo}, {Karbevska}, {K{\'o}sp{\'a}l}, {Kruszy{\'n}ska}, {Kun}, {Lanza}, {Leccia}, {Marconi}, {Messina}, {Molinaro}, {Moln{\'a}r}, {Muraveva}, {Musella}, {Nagy}, {Pagano}, {Palaversa}, {Plachy}, {Pr{\v{s}}a}, {Rybicki}, {Shahaf}, {Szabados}, {Szegedi-Elek}, {Trabucchi}, {Barblan}, {Grenon}, {Roelens}, \& {S{\"u}veges}}]{Eyer2023}
{Eyer}, L., {Audard}, M., {Holl}, B., {et~al.} 2023, \aap, 674, A13

\bibitem[{{Feast}(1997)}]{Feast1997b}
{Feast}, M.~W. 1997, \mnras, 285, 339

\bibitem[{{Fryer} \& {Diehl}(2008)}]{Fryer2008}
{Fryer}, C.~L., \& {Diehl}, S. 2008, in Astronomical Society of the Pacific Conference Series, Vol. 391, Hydrogen-Deficient Stars, ed. A.~{Werner} \& T.~{Rauch}, 335

\bibitem[{{Fuenmayor}(1981)}]{Fuenmayor1981}
{Fuenmayor}, F.~J. 1981, \rmxaa, 6, 83

\bibitem[{{Garc{\'\i}a-Hern{\'a}ndez} {et~al.}(2023){Garc{\'\i}a-Hern{\'a}ndez}, {Rao}, {Lambert}, {Eriksson}, {Reddy}, \& {Masseron}}]{Garcia-Hernandez2023}
{Garc{\'\i}a-Hern{\'a}ndez}, D.~A., {Rao}, N.~K., {Lambert}, D.~L., {et~al.} 2023, \apj, 948, 15

\bibitem[{{Gautschy}(2023)}]{Gautschy2023}
{Gautschy}, A. 2023, arXiv e-prints, arXiv:2312.14693

\bibitem[{{Gautschy-Loidl} {et~al.}(2004){Gautschy-Loidl}, {H{\"o}fner}, {J{\o}rgensen}, \& {Hron}}]{GautschyLoidl2004}
{Gautschy-Loidl}, R., {H{\"o}fner}, S., {J{\o}rgensen}, U.~G., \& {Hron}, J. 2004, \aap, 422, 289

\bibitem[{{Gonneau} {et~al.}(2016){Gonneau}, {Lan{\c{c}}on}, {Trager}, {Aringer}, {Lyubenova}, {Nowotny}, {Peletier}, {Prugniel}, {Chen}, {Dries}, {Choudhury}, {Falc{\'o}n-Barroso}, {Koleva}, {Meneses-Goytia}, {S{\'a}nchez-Bl{\'a}zquez}, \& {Vazdekis}}]{Gonneau2016}
{Gonneau}, A., {Lan{\c{c}}on}, A., {Trager}, S.~C., {et~al.} 2016, \aap, 589, A36

\bibitem[{{Gray} \& {Corbally}(2009)}]{Gray2009}
{Gray}, R.~O., \& {Corbally}, Christopher, J. 2009, {Stellar Spectral Classification}

\bibitem[{{Green} {et~al.}(2019){Green}, {Schlafly}, {Zucker}, {Speagle}, \& {Finkbeiner}}]{Green2019}
{Green}, G.~M., {Schlafly}, E., {Zucker}, C., {Speagle}, J.~S., \& {Finkbeiner}, D. 2019, \apj, 887, 93

\bibitem[{{Han}(1998)}]{Han98}
{Han}, Z. 1998, \mnras, 296, 1019

\bibitem[{{Herter} {et~al.}(2008){Herter}, {Henderson}, {Wilson}, {Matthews}, {Rahmer}, {Bonati}, {Muirhead}, {Adams}, {Lloyd}, {Skrutskie}, {Moon}, {Parshley}, {Nelson}, {Martinache}, \& {Gull}}]{Herter2008}
{Herter}, T.~L., {Henderson}, C.~P., {Wilson}, J.~C., {et~al.} 2008, in Society of Photo-Optical Instrumentation Engineers (SPIE) Conference Series, Vol. 7014, \procspie, 70140X

\bibitem[{{Karakas} {et~al.}(2015){Karakas}, {Ruiter}, \& {Hampel}}]{Karakas2015}
{Karakas}, A.~I., {Ruiter}, A.~J., \& {Hampel}, M. 2015, \apj, 809, 184

\bibitem[{{Karambelkar} {et~al.}(2022){Karambelkar}, {Kasliwal}, {Tisserand}, {Clayton}, {Crawford}, {Anand}, {Geballe}, \& {Montiel}}]{Karambelkar2022}
{Karambelkar}, V., {Kasliwal}, M.~M., {Tisserand}, P., {et~al.} 2022, \aap, 667, A84

\bibitem[{{Karambelkar} {et~al.}(2021){Karambelkar}, {Kasliwal}, {Tisserand}, {De}, {Anand}, {Ashley}, {Delacroix}, {Hankins}, {Jencson}, {Lau}, {McKenna}, {Moore}, {Ofek}, {Smith}, {Soria}, {Soon}, {Tinyanont}, {Travouillon}, \& {Yao}}]{Karambelkar2021}
{Karambelkar}, V.~R., {Kasliwal}, M.~M., {Tisserand}, P., {et~al.} 2021, \apj, 910, 132

\bibitem[{{Lamberts} {et~al.}(2019){Lamberts}, {Blunt}, {Littenberg}, {Garrison-Kimmel}, {Kupfer}, \& {Sanderson}}]{Lamberts2019}
{Lamberts}, A., {Blunt}, S., {Littenberg}, T.~B., {et~al.} 2019, \mnras, 490, 5888

\bibitem[{{Lawson} \& {Cottrell}(1997)}]{Lawson1997}
{Lawson}, W.~A., \& {Cottrell}, P.~L. 1997, \mnras, 285, 266

\bibitem[{{Lawson} {et~al.}(1990){Lawson}, {Cottrelll}, {Kilmartin}, \& {Gilmore}}]{Lawson1990}
{Lawson}, W.~A., {Cottrelll}, P.~L., {Kilmartin}, P.~M., \& {Gilmore}, A.~C. 1990, \mnras, 247, 91

\bibitem[{{Lawson} \& {Kilkenny}(1996)}]{Lawson1996}
{Lawson}, W.~A., \& {Kilkenny}, D. 1996, in Astronomical Society of the Pacific Conference Series, Vol.~96, Hydrogen Deficient Stars, ed. C.~S. {Jeffery} \& U.~{Heber}, 349

\bibitem[{{Lee}(2015)}]{Lee2015}
{Lee}, C.~H. 2015, \aap, 575, A2

\bibitem[{Lee {et~al.}(2020)Lee, Matheson, Soraisam, Narayan, Saha, Stubens, \& Wolf}]{Lee2020}
Lee, C.-H., Matheson, T., Soraisam, M., {et~al.} 2020, The Astronomical Journal, 159, 61

\bibitem[{{Lee} {et~al.}(1947){Lee}, {Gore}, \& {Bartlett}}]{Lee1947}
{Lee}, O.~J., {Gore}, G., \& {Bartlett}, T.~J. 1947, Annals of the Dearborn Observatory, 5, 287

\bibitem[{{Lourie} {et~al.}(2020){Lourie}, {Baker}, {Burruss}, {Egan}, {F{\.z}r{\'e}sz}, {Frostig}, {Garcia-Zych}, {Ganciu}, {Haworth}, {Hinrichsen}, {Kasliwal}, {Karambelkar}, {Malonis}, {Simcoe}, \& {Zolkower}}]{Lourie2020}
{Lourie}, N.~P., {Baker}, J.~W., {Burruss}, R.~S., {et~al.} 2020, in Society of Photo-Optical Instrumentation Engineers (SPIE) Conference Series, Vol. 11447, Ground-based and Airborne Instrumentation for Astronomy VIII, ed. C.~J. {Evans}, J.~J. {Bryant}, \& K.~{Motohara}, 114479K

\bibitem[{{Maehara} \& {Soyano}(1987)}]{Maehara1987}
{Maehara}, H., \& {Soyano}, T. 1987, Annals of the Tokyo Astronomical Observatory, 21, 293

\bibitem[{{Ma{\'\i}z Apell{\'a}niz} {et~al.}(2023){Ma{\'\i}z Apell{\'a}niz}, {Holgado}, {Pantaleoni Gonz{\'a}lez}, \& {Caballero}}]{Apellaniz2023}
{Ma{\'\i}z Apell{\'a}niz}, J., {Holgado}, G., {Pantaleoni Gonz{\'a}lez}, M., \& {Caballero}, J.~A. 2023, \aap, 677, A137

\bibitem[{{Nassau} \& {Blanco}(1957)}]{Nassau1957}
{Nassau}, J.~J., \& {Blanco}, V.~M. 1957, \apj, 125, 195

\bibitem[{{Otero} {et~al.}(2014){Otero}, {H{\"u}mmerich}, {Bernhard}, \& {Sozynski}}]{Otero2014}
{Otero}, S., {H{\"u}mmerich}, S., {Bernhard}, K., \& {Sozynski}, I. 2014, Journal of the American Association of Variable Star Observers (JAAVSO), 42, 13

\bibitem[{{Percy}(2023)}]{Percy2023}
{Percy}, J.~R. 2023, \jaavso, 51, 64

\bibitem[{{Putney}(1997)}]{Putney1997}
{Putney}, A. 1997, \apjs, 112, 527

\bibitem[{{Rayner} {et~al.}(2009){Rayner}, {Cushing}, \& {Vacca}}]{Rayner2009}
{Rayner}, J.~T., {Cushing}, M.~C., \& {Vacca}, W.~D. 2009, \apjs, 185, 289

\bibitem[{{Rayner} {et~al.}(2003){Rayner}, {Toomey}, {Onaka}, {Denault}, {Stahlberger}, {Vacca}, {Cushing}, \& {Wang}}]{Rayner2003}
{Rayner}, J.~T., {Toomey}, D.~W., {Onaka}, P.~M., {et~al.} 2003, \pasp, 115, 362

\bibitem[{{Ricker} {et~al.}(2015){Ricker}, {Winn}, {Vanderspek}, {Latham}, {Bakos}, {Bean}, {Berta-Thompson}, {Brown}, {Buchhave}, {Butler}, {Butler}, {Chaplin}, {Charbonneau}, {Christensen-Dalsgaard}, {Clampin}, {Deming}, {Doty}, {De Lee}, {Dressing}, {Dunham}, {Endl}, {Fressin}, {Ge}, {Henning}, {Holman}, {Howard}, {Ida}, {Jenkins}, {Jernigan}, {Johnson}, {Kaltenegger}, {Kawai}, {Kjeldsen}, {Laughlin}, {Levine}, {Lin}, {Lissauer}, {MacQueen}, {Marcy}, {McCullough}, {Morton}, {Narita}, {Paegert}, {Palle}, {Pepe}, {Pepper}, {Quirrenbach}, {Rinehart}, {Sasselov}, {Sato}, {Seager}, {Sozzetti}, {Stassun}, {Sullivan}, {Szentgyorgyi}, {Torres}, {Udry}, \& {Villasenor}}]{Ricker2015}
{Ricker}, G.~R., {Winn}, J.~N., {Vanderspek}, R., {et~al.} 2015, Journal of Astronomical Telescopes, Instruments, and Systems, 1, 014003

\bibitem[{{Saio}(2008)}]{Saio2008}
{Saio}, H. 2008, in Astronomical Society of the Pacific Conference Series, Vol. 391, Hydrogen-Deficient Stars, ed. A.~{Werner} \& T.~{Rauch}, 69

\bibitem[{{Schwab}(2019)}]{Schwab2019}
{Schwab}, J. 2019, \apj, 885, 27

\bibitem[{{Shields} {et~al.}(2019){Shields}, {Jayasinghe}, {Stanek}, {Kochanek}, {Shappee}, {Holoien}, {Thompson}, {Prieto}, \& {Dong}}]{Shields2019}
{Shields}, J.~V., {Jayasinghe}, T., {Stanek}, K.~Z., {et~al.} 2019, \mnras, 483, 4470

\bibitem[{{Smith} {et~al.}(2020){Smith}, {Smartt}, {Young}, {Tonry}, {Denneau}, {Flewelling}, {Heinze}, {Weiland}, {Stalder}, {Rest}, {Stubbs}, {Anderson}, {Chen}, {Clark}, {Do}, {F{\"o}rster}, {Fulton}, {Gillanders}, {McBrien}, {O'Neill}, {Srivastav}, \& {Wright}}]{Smith2020atlas}
{Smith}, K.~W., {Smartt}, S.~J., {Young}, D.~R., {et~al.} 2020, \pasp, 132, 085002

\bibitem[{{Soszy{\'n}ski} {et~al.}(2009){Soszy{\'n}ski}, {Udalski}, {Szyma{\'n}ski}, {Kubiak}, {Pietrzy{\'n}ski}, {Wyrzykowski}, {Szewczyk}, {Ulaczyk}, \& {Poleski}}]{Soszynski2009}
{Soszy{\'n}ski}, I., {Udalski}, A., {Szyma{\'n}ski}, M.~K., {et~al.} 2009, \actaa, 59, 335

\bibitem[{{Soyano} \& {Maehara}(1991)}]{Soyano1991}
{Soyano}, T., \& {Maehara}, H. 1991, Publications of the National Astronomical Observatory of Japan, 2, 203

\bibitem[{{Stephenson}(1973)}]{Stephenson1973}
{Stephenson}, C.~B. 1973, Publications of the Warner \& Swasey Observatory

\bibitem[{{Tang} {et~al.}(2013){Tang}, {Cao}, {Bildsten}, {Nugent}, {Bellm}, {Kulkarni}, {Laher}, {Levitan}, {Masci}, {Ofek}, {Prince}, {Sesar}, \& {Surace}}]{Tang2013}
{Tang}, S., {Cao}, Y., {Bildsten}, L., {et~al.} 2013, \apjl, 767, L23

\bibitem[{{Tisserand}(2012)}]{Tisserand2012}
{Tisserand}, P. 2012, \aap, 539, A51

\bibitem[{{Tisserand} {et~al.}(2013){Tisserand}, {Clayton}, {Welch}, {Pilecki}, {Wyrzykowski}, \& {Kilkenny}}]{Tisserand2013}
{Tisserand}, P., {Clayton}, G.~C., {Welch}, D.~L., {et~al.} 2013, \aap, 551, A77

\bibitem[{{Tisserand} {et~al.}(2024{\natexlab{a}}){Tisserand}, {Crawford}, {Soon}, {Clayton}, {Ruiter}, \& {Seitenzahl}}]{Tisserand2023b}
{Tisserand}, P., {Crawford}, C.~L., {Soon}, J., {et~al.} 2024{\natexlab{a}}, \aap, 684, A131

\bibitem[{{Tisserand} {et~al.}(2024{\natexlab{b}}){Tisserand}, {Crawford}, {Soon}, {Clayton}, {Ruiter}, \& {Seitenzahl}}]{Tisserand2023a}
---. 2024{\natexlab{b}}, \aap, 684, A130

\bibitem[{{Tisserand} {et~al.}(2004){Tisserand}, {Marquette}, {Beaulieu}, {de Laverny}, {Lesquoy}, {Milsztajn}, {Afonso}, {Albert}, {Andersen}, {Ansari}, {Aubourg}, {Bareyre}, {Bauer}, {Blanc}, {Charlot}, {Coutures}, {Derue}, {Ferlet}, {Fouqu{\'e}}, {Glicenstein}, {Goldman}, {Gould}, {Graff}, {Gros}, {Haissinski}, {Hamadache}, {de Kat}, {Lasserre}, {Le Guillou}, {Loup}, {Magneville}, {Mansoux}, {Maurice}, {Maury}, {Moniez}, {Palanque-Delabrouille}, {Perdereau}, {Pr{\'e}vot}, {Rahal}, {Regnault}, {Rich}, {Spiro}, {Vidal-Madjar}, {Vigroux}, \& {Zylberajch}}]{Tisserand2004}
{Tisserand}, P., {Marquette}, J.~B., {Beaulieu}, J.~P., {et~al.} 2004, \aap, 424, 245

\bibitem[{{Tisserand} {et~al.}(2008){Tisserand}, {Marquette}, {Wood}, {Lesquoy}, {Beaulieu}, {Milsztajn}, {Hamadache}, {Afonso}, {Albert}, {Andersen}, {Ansari}, {Aubourg}, {Bareyre}, {Charlot}, {Coutures}, {Ferlet}, {Fouqu{\'e}}, {Glicenstein}, {Goldman}, {Gould}, {Gros}, {Haissinski}, {de Kat}, {Le Guillou}, {Loup}, {Magneville}, {Maurice}, {Maury}, {Moniez}, {Palanque-Delabrouille}, {Perdereau}, {Rahal}, {Rich}, {Spiro}, {Vidal-Madjar}, \& {Zylberajch}}]{Tisserand2008}
{Tisserand}, P., {Marquette}, J.~B., {Wood}, P.~R., {et~al.} 2008, \aap, 481, 673

\bibitem[{{Tisserand} {et~al.}(2009){Tisserand}, {Wood}, {Marquette}, {Afonso}, {Albert}, {Andersen}, {Ansari}, {Aubourg}, {Bareyre}, {Beaulieu}, {Charlot}, {Coutures}, {Ferlet}, {Fouqu{\'e}}, {Glicenstein}, {Goldman}, {Gould}, {Gros}, {de Kat}, {Lesquoy}, {Loup}, {Magneville}, {Maurice}, {Maury}, {Milsztajn}, {Moniez}, {Palanque-Delabrouille}, {Perdereau}, {Rich}, {Schwemling}, {Spiro}, \& {Vidal-Madjar}}]{Tisserand2009}
{Tisserand}, P., {Wood}, P.~R., {Marquette}, J.~B., {et~al.} 2009, \aap, 501, 985

\bibitem[{{Tisserand} {et~al.}(2011){Tisserand}, {Wyrzykowski}, {Wood}, {Udalski}, {Szyma{\'n}ski}, {Kubiak}, {Pietrzy{\'n}ski}, {Soszy{\'n}ski}, {Szewczyk}, {Ulaczyk}, \& {Poleski}}]{Tisserand2011}
{Tisserand}, P., {Wyrzykowski}, L., {Wood}, P.~R., {et~al.} 2011, \aap, 529, A118

\bibitem[{{Tisserand} {et~al.}(2020){Tisserand}, {Clayton}, {Bessell}, {Welch}, {Kamath}, {Wood}, {Wils}, {Wyrzykowski}, {Mr{\'o}z}, \& {Udalski}}]{Tisserand2020}
{Tisserand}, P., {Clayton}, G.~C., {Bessell}, M.~S., {et~al.} 2020, \aap, 635, A14

\bibitem[{{Tisserand} {et~al.}(2022){Tisserand}, {Crawford}, {Clayton}, {Ruiter}, {Karambelkar}, {Bessell}, {Seitenzahl}, {Kasliwal}, {Soon}, \& {Travouillon}}]{Tisserand2022}
{Tisserand}, P., {Crawford}, C.~L., {Clayton}, G.~C., {et~al.} 2022, \aap, 667, A83

\bibitem[{{Tonry} {et~al.}(2018){Tonry}, {Denneau}, {Heinze}, {Stalder}, {Smith}, {Smartt}, {Stubbs}, {Weiland}, \& {Rest}}]{Tonry2018}
{Tonry}, J.~L., {Denneau}, L., {Heinze}, A.~N., {et~al.} 2018, \pasp, 130, 064505

\bibitem[{{Vacca} {et~al.}(2003){Vacca}, {Cushing}, \& {Rayner}}]{Vacca2003}
{Vacca}, W.~D., {Cushing}, M.~C., \& {Rayner}, J.~T. 2003, \pasp, 115, 389

\bibitem[{von Neumann(1941)}]{vonNeumann1941}
von Neumann, J. 1941, The Annals of Mathematical Statistics, 12, 367

\bibitem[{{Warner}(1967)}]{Warner1967}
{Warner}, B. 1967, \mnras, 137, 119

\bibitem[{{Webbink}(1984)}]{Webbink84}
{Webbink}, R.~F. 1984, \apj, 277, 355

\bibitem[{{Wong} \& {Bildsten}(2024)}]{Wong2024}
{Wong}, T. L.~S., \& {Bildsten}, L. 2024, \apj, 962, 20

\bibitem[{{Zaniewski} {et~al.}(2005){Zaniewski}, {Clayton}, {Welch}, {Gordon}, {Minniti}, \& {Cook}}]{Zaniewski2005}
{Zaniewski}, A., {Clayton}, G.~C., {Welch}, D.~L., {et~al.} 2005, \aj, 130, 2293

\bibitem[{{Za{\v{c}}s} {et~al.}(2007){Za{\v{c}}s}, {Mondal}, {Chen}, {Pugach}, {Musaev}, \& {Alksnis}}]{Zacs2007}
{Za{\v{c}}s}, L., {Mondal}, S., {Chen}, W.~P., {et~al.} 2007, \aap, 472, 247

\end{thebibliography}
\bibliographystyle{apj}

\label{lastpage}
\end{document}